\documentclass[fleqn,usenatbib]{mnras}

\usepackage{newtxtext,newtxmath}
\usepackage[T1]{fontenc}
\usepackage{graphicx}	
\usepackage{amsmath}	
\usepackage[usenames,dvipsnames]{xcolor}
\usepackage{hyperref}

\newcommand{\pd}{\partial}
\newcommand{\erf}{\text{erf}}
\newcommand{\bigO}{\mathcal{O}}
\newcommand{\bigM}{\mathcal{M}}
\newcommand{\eref}[1]{Eq.~(\ref{#1})}
\newcommand{\pCAeref}[1]{Eqs.~(\ref{#1}}
\newcommand{\pCBeref}[1]{, \ref{#1}}
\newcommand{\pCCeref}[1]{, \ref{#1})}
\newcommand{\nsub}{n_\text{sub}}
\newcommand{\msub}{M_\text{sub}}
\newcommand{\csub}{c_\text{sub}}
\newcommand{\alphasub}{{\alpha_\text{sub}}}
\newcommand{\betasub}{{\beta_\text{sub}}}
\newcommand{\redz}{\mathrm{z}}
\newcommand{\tage}{t_\text{age}}
\newcommand{\tcos}{t_\text{cos}}
\newcommand{\Mp}{M_\text{p}}
\newcommand{\vp}{v_\text{p}}
\newcommand{\Mh}{M_\text{h}}
\newcommand{\Mgra}{M_\text{gra}}
\newcommand{\sigmaeff}{\sigma_\text{eff}}
\newcommand{\sigmah}{\sigma_\text{h}}
\newcommand{\Xeff}{X_\text{eff}}
\newcommand{\bigG}{\mathbb{G}}
\newcommand{\bigE}{\mathbb{E}}
\newcommand{\bigD}{\mathcal{D}}
\newcommand{\rt}{r_\text{tid}}
\newcommand{\rs}{r_\text{s}}
\newcommand{\ft}{f_\text{tid}}
\newcommand{\rc}{r_\text{c}}
\newcommand{\rhoc}{\rho_\text{c}}
\newcommand{\zmax}{z_\text{max}}
\newcommand{\vc}{v_\text{cir}}
\newcommand{\vbf}{\boldsymbol{v}}
\newcommand{\bigH}{\mathcal{H}}
\newcommand{\rhobgeff}{\rho_\text{bg}^\text{eff}}
\newcommand{\rhoh}{\rho_\text{h}}
\newcommand{\rhod}{\rho_\text{d}}
\newcommand{\<}{\langle}
\renewcommand{\>}{\rangle}
\newcommand{\eee}{\equiv}

\defcitealias{Church2019MNRAS}{C19}

\title[Ultralight DM and MW disc heating]
      {Can ultralight dark matter explain the age\textendash velocity dispersion relation of the Milky Way disc: A revised and improved treatment}

\author[B. T. Chiang, J. P. Ostriker and H.-Y. Schive]{
Barry T. Chiang$^{1,2}$\thanks{E-mail: btjc2@cantab.ac.uk},
Jeremiah P. Ostriker$^{1,3}$\thanks{Columbia Contact E-mail: jpo@astro.columbia.edu 
	\newline
	Princeton Contact E-mail: ostriker@princeton.edu}, and
Hsi-Yu Schive$^{2,4,5,6}$
\vspace*{8pt}
\\
$^{1}$Department of Astronomy, Columbia University, New York, NY 10027, USA.\\
$^{2}$Institute of Astrophysics, National Taiwan University, Taipei 10617, Taiwan.\\
$^{3}$Department of Astrophysical Sciences, Princeton University, 4 Ivy Lane, Princeton, NJ, 08544, USA.\\
$^{4}$Department of Physics, National Taiwan University, Taipei 10617, Taiwan.\\
$^{5}$Center for Theoretical Physics, National Taiwan University, Taipei 10617, Taiwan.\\
$^{6}$Physics Division, National Center for Theoretical Sciences, Taipei 10617, Taiwan.}

\date{Accepted XXX. Received YYY; in original form ZZZ}

\pubyear{2022}

\begin{document}

\label{firstpage}
\pagerange{\pageref{firstpage}--\pageref{lastpage}}
\maketitle


\begin{abstract}
	Ultralight axion-like particles $m_a \sim 10^{-22}$~eV, or Fuzzy Dark Matter (FDM), behave comparably to cold dark matter (CDM) on cosmological scales and exhibit a kpc-size de Broglie wavelength capable of alleviating established (sub-)galactic-scale problems of CDM. Substructures inside an FDM halo incur gravitational potential perturbations, resulting in stellar heating sufficient to account for the Galactic disc thickening over a Hubble time, as first demonstrated by Church et al. We present a more sophisticated treatment that incorporates the full baryon and dark matter distributions of the Milky Way and adopts stellar disc kinematics inferred from recent \textit{Gaia}, APOGEE, and LAMOST surveys. Ubiquitous density granulation and subhalo passages respectively drive inner disc thickening and flaring of the outer disc, resulting in an observationally consistent `U-shaped' disc vertical velocity dispersion profile with the global minimum located near the solar radius. The observed age\textendash velocity dispersion relation in the solar vicinity can be explained by the FDM-substructure-induced heating and places an exclusion bound $m_a \gtrsim 0.4\times10^{-22}$ eV. We assess non-trivial uncertainties in the empirical core-halo relation, FDM subhalo mass function and tidal stripping, and stellar heating estimate. The mass range $m_a\simeq 0.5$\textendash$0.7\times10^{-22}$~eV favoured by the observed thick disc kinematics is in tension with several exclusion bounds inferred from dwarf density profiles, stellar streams, and Milky Way satellite populations, which could be significantly relaxed due to the aforesaid uncertainties. Additionally, strongly anisotropic heating could help explain the formation of ultra-thin disc galaxies.
\end{abstract}



\section{Introduction}\label{sec:intro}

The Milky Way (MW) stellar disc comprises various kinematically distinct components \citep[e.g.][]{Bovy2012ApJ751,Bovy2012ApJ753,Miglio2021A&A645A} that have scale heights increasing with the characteristic ages \citep{Bovy2016ApJ823, Ferguson2017ApJ843, Lu2022MNRAS512}. The oldest populations with age $\gtrsim 8$ Gyr, the \textit{thick disc}, exhibit high $[\alpha/\text{Fe}]$ abundance enriched by early-time Type
II supernovae and substantially higher velocity dispersion \citep[e.g.][]{Mackereth2019MNRAS,Ting2019ApJ878,Sharma2021MNRAS506}, relative to the \textit{thin disc} dominated by kinematically cold, low-$[\alpha/\text{Fe}]$ populations younger than 8 Gyr \citep[e.g.][]{Bland-Hawthorn2016ARA&A54,Duong2018MNRAS476}. Comparing the typical scale height of the thick disc $\simeq0.7$~kpc with $\simeq 0.3$~kpc of the thin disc \citep{Bovy2012ApJ753,Bland-Hawthorn2016ARA&A54} indicates significant dynamical heating over a Hubble time. This morphological evolution and the observed correlation between the disc stellar age and random motions in the solar vicinity \citep[e.g.][]{Aumer2009MNRAS397,Miglio2021A&A645A} provide a unique probe into the formation history of the Galaxy.

Amongst a number of proposed and qualitatively different heating mechanisms, there has yet to be an entirely satisfactory explanation for the origin of the thick disc. Interactions with giant molecular clouds \citep[e.g.][]{Spitzer1953ApJ118,Lacey1984MNRAS208}, albeit capable of reproducing the age\textendash velocity relation of the inner disc, cannot explain the formation of the Galactic thick disc nor the observed flaring of low-$[\alpha/\text{Fe}]$ populations in the outer disc  \citep{Aumer2016MNRAS459,Mackereth2019MNRAS,Sharma2021MNRAS506}. Transient spiral modes \citep[e.g.][]{Barbanis1967ApJ150,Lynden-Bell1972MNRAS157} fail to sufficiently increase the vertical dispersion in MW-like disc galaxies \citep{Sellwood2013ApJ769L, MartinezMedina2015ApJ802}. Similarly, \citet{Grand2016MNRAS459} demonstrate with high-resolution cosmological simulations that the dynamical heating induced by spiral structures is subleading to bar corotation \citep{Minchev2010ApJ722} and perturbations from subhaloes heavier than $\bigO(10^{10})$ M$_\odot$ \citep{Villalobos2008j1365}. Small satellite accretion over cosmological epochs could lead to non-trivial heating only in the outer disc region \citep[e.g.][]{Toth1992ApJ389, Quinn1993ApJ403, Moetazedian2016MNRAS459}.

Radial migration, another secular process of great theoretical interest \citep[e.g.][]{Sellwood2002MNRAS336,Minchev2010ApJ722}, has amassed considerable observational support. Intermediate-old, super-metal-rich populations present in the solar vicinity bear a chemical evolution dissimilar to that of the local thin disc stars \citep{Kordopatis2015MNRAS447,Anders2018AA619A,Hayden2020MNRAS493}, and are found to have migrated radially outwards by about 4\textendash6 kpc over the formation timescale of the thin disc $\simeq 8$ Gyr \citep{Minchev2013AA558A,Frankel2018ApJ865, Frankel2019ApJ884F,Miglio2021A&A645A}. These high-$[\alpha/\text{Fe}]$, kinematically hot stars were first formed in the inner disc region, and migrate outwards to make up the high-$\sigma_z$ tail of the age\textendash velocity dispersion relationship in the solar neighbourhood \citep{Sharma2021MNRAS506,Lu2022MNRAS512}. However, \citet{Minchev2016AN337} concludes from disc chemodynamics that radial migration per se does not lead to effective $z$-direction stellar heating.

Recent large-scale spectroscopic surveys such as \textit{Gaia} \citep{GaiaCollaboration2016A&A595A}, APOGEE \citep{Majewski2017AJ154}, \textit{Kepler} \citep{Borucki2010Sci327}, LAMOST \citep{Zhao2012LAMOST}, and GALAH \citep{Martell2017MNRAS465} have made possible increasingly precise measurements of the disc vertical velocity dispersion profile $\sigma_z(R)$, age information, and spatially resolved chemical structures. High-resolution cosmological simulations further put  to test many of the proposed secular mechanisms, with the aim of clarifying the Galactic disc formation history \citep[e.g.][]{Buck2020MNRAS491, Agertz2021MNRAS503, Okalidis2022MNRAStmp}. In view of these recent developments, here we consider in detail a disc formation scenario driven by dark sector physics first investigated in \citet{Church2019MNRAS}, henceforth \citetalias{Church2019MNRAS}. This alternative disc heating mechanism arises from the gravitational potential fluctuations caused by ultralight dark matter (DM) substructures.

In the presence of various small-scale astrophysical observables possibly incompatible with the standard cold dark matter (CDM) predictions (see \citealp{Bullock:2017xww} for a review), \mbox{sub-eV} bosonic DM represents an alternative, broad, and theoretically well-motivated class of models \citep[e.g.][]{Svrcek:2006yi,Marsh:2015xka,Hui:2016ltb,Halverson:2017deq} with distinctive wave signatures that can potentially be probed by terrestrial experiments \citep[e.g.][]{Graham:2015ouw,Derevianko:2016vpm,Centers:2019dyn,Roussy:2020ily,Lisanti:2021vij} or galactic-scale observations \citep[e.g.][]{Niemeyer:2019aqm,AEDGE:2019nxb,Fedderke:2019ajk,Reynolds2020ApJ890,Hui:2021tkt}. In the ultralight regime $\sim 10^{-22}$~eV, Fuzzy Dark Matter (FDM), first coined by \citet{Hu:2000ke}, features a kpc/sub-kpc-size de Broglie wavelength and a wide range of astrophysical implications that have been actively explored via self-consistent numerical simulations \citep[e.g.][]{Schive:2014dra, Schive:2014hza,Schive:2019rrw, Schwabe:2016rze,Schwabe:2020eac, Mocz:2017wlg, Mocz:2019pyf, Chan2018MNRAS, Veltmaat:2018dfz, Veltmaat:2019hou, Chowdhury2021ApJ, May:2021wwp,May2022arXiv220914886M, Chiang:2021uvt, Chan:2021bja}. The ubiquitous quantum fluctuations and orbiting subhaloes within a host FDM halo could cause significant dynamical heating in both the inner and outer Galactic disc regions \citepalias{Church2019MNRAS}. These locally fluctuating granules weighting $\gtrsim 10^5$~M$_\odot$ and with a velocity of  $\simeq 100$ km s$^{-1}$ in an MW-sized halo are capable of accelerating \mbox{stars and possibly explaining the origin of the thick disc.}

\citetalias{Church2019MNRAS} made a first, rough, preliminary estimate of these effects and found that with the preferred particle mass of $m_a \simeq 0.7\times10^{-22}$~eV, the amount of disc heating in the solar vicinity can be obtained with the observationally correct relation between scale height and stellar age. The purpose of this paper is to make a substantially more accurate calculation of the FDM heating process and carefully assess the resulting predictions across the entire Galactic disc with recent kinematic measurements. 

In Sec. \ref{sec:Milky_Way_Profiles}, we present a self-consistent treatment of the Galactic disc that accounts for the background matter distribution of the MW. Disc heating due to interactions with FDM substructures is analytically quantified in Sec. \ref{sec:Dynamical_Evolution_Disc}. The predicted age\textendash velocity dispersion relations in the FDM and CDM paradigms as well as the resultant FDM particle mass constraint are discussed in Sec.~\ref{ssec:FDM_Heating_Predictions}; the dynamical impact of the Gaia\textendash Enceladus and Sagittarius (Sgr) dwarf mergers is also examined. In Sec.~\ref{ssec:Sources_of_Error} we inspect non-trivial modelling uncertainties in the empirical core-halo relation, FDM subhalo mass function, and FDM substructure-driven stellar heating estimate and comment on the main differences between our current framework and that of \citetalias{Church2019MNRAS}. Implications on the robustness of relevant FDM constraints are remarked in Sec. \ref{ssec:Literature_Comparison}. Section \ref{ssec:Anisotropic_Heating} explores a possible FDM-induced formation mechanism of ultra-thin disc galaxies. In Sec. \ref{sec:Conclusions}, we conclude. The companion paper (Yang et al., in preparation), complementary to the present analytical calculations, explores the FDM disc heating mechanism with fully \mbox{self-consistent dynamical simulations.}
  	
In this manuscript, cylindrical coordinates $(R,z)$ denote the axial distance along and vertical height  from the mid-plane in the Galactocentric frame; $r \eee \sqrt{R^2+z^2}$ gives the radial distance from the Galactic centre. Cosmological redshift is denoted by $\redz$. The age of a stellar population $\tage$ should not be confused with the cosmic time $\tcos$ (i.e. $\tcos \simeq 13.8$ Gyr at redshift zero).

\section{The Milky Way Modelling}
\label{sec:Milky_Way_Profiles}

We derive in Sec. \ref{ssec:Disc_Profile_Modelling} the analytical disc density profiles, scale heights, and gravitational potentials when the vertical gravity is dominantly sourced by either the local disc or other non-disc/non-local background mass distributions. The baryon mass infall history and inside-out formation of the Galactic disc are considered in Sec.~\ref{ssec:Disc_Mass_Accretion}. We reconstruct a realistic MW mass background potential and review the relevant observational constraints in Sec. \ref{ssec:MW_Background_Disc_sigma_z}.


\subsection{Self-Consistent Stellar Disc Profile Modelling}\label{ssec:Disc_Profile_Modelling}

The Galactic thick and thin disc components are conventionally associated with distinct scale lengths, heights, kinematics, and formation histories \citep[e.g.][]{Bland-Hawthorn2016ARA&A54}. However, the exact thick-to-thin disc mass ratio is still debated, with values ranging from 10\%\textendash25\% \citep[e.g.][]{Robin:2003uus,McMillan2011MNRAS414,Barros2016A&A593A}, $33\%$ \citep{Kubryk2015A&A580A}, to about $50\%$ \citep[e.g.][]{Fuhrmann2012MNRAS,Haywood2013A&A560A,Snaith2014ApJ781L}. This dichotomous picture, rests upon the defining bi-modality in $[\alpha/\text{Fe}]$ abundance, is further complicated by the high-resolution spectroscopic data from SEGUE \citep{Yanny2009AJ137}, LAMOST \citep{Zhao2012LAMOST}, and APOGEE \citep{Majewski2017AJ154} that appear to favour a multi-component decomposition comprising mono-abundance sub-populations \citep{Bovy2012ApJ751,Bovy2012ApJ753, Bovy2016ApJ823,Yu2021ApJ912}. In light of these non-trivial modelling uncertainties, we treat the Galactic disc populations collectively as a single gravitationally bound structure that follows an inside-out formation pattern with an observationally constrained mass accretion history, as detailed in Sec. \ref{ssec:Disc_Mass_Accretion}.

The projected, axisymmetric stellar disc density distribution empirically follows an exponential surface density profile
\begin{eqnarray}\label{eqn:Expo_Surface_Den}
	\Sigma(R) = \int_{-\infty}^\infty dz \rhod(R,z) = \Sigma_0 e^{-R/R_0},
\end{eqnarray}
where $\rhod(R,z)$ denotes the 3D disc density profile and $\Sigma(R_\odot) = 68$~M$_\odot$pc$^{-2}$ in the solar neighbourhood $R_\odot = 8.3$~kpc \citep{Bovy:2013raa,Zhang2013ApJ772}. Measurements of the MW disc scale length still exhibit large spread \citep{Bland-Hawthorn2016ARA&A54} and are further complicated by the debated disc substructure decomposition \citep[e.g.][]{Bovy2012ApJ751,Lian2020MNRAS497}. Following the treatment in \citetalias{Church2019MNRAS}, we adopt $R_0 = 3.2$~kpc that reasonably reproduces the component-wise rotation curve in \citet{Kafle:2014xfa}, giving a disc total mass $M_\text{d} = 5.9\times 10^{10}$ M$_\odot$ consistent with, for example, \citet{Licquia2015ApJ806} and \citet{Bland-Hawthorn2016ARA&A54}.

The vertical velocity dispersion $\sigma_z$ of the thick disc shows weak or no $z$-dependence across $3$~kpc~$\lesssim R\lesssim 20$ kpc and $|z| \lesssim 3$ kpc \citep{Mackereth2019MNRAS, Sharma2021MNRAS506}. This observed isothermality $\pd \sigma_z/\pd z \simeq 0$ indicates the Galactic disc stars are kinematically decoupled along the radial and vertical directions to a good approximation, giving the following disc potential decomposition \citep{Lacey1985ApJ,Binney&Tremaine2008}
\begin{eqnarray}\label{eqn:Disc_Gra_Potential} 
	\begin{cases}
		\phi_\text{d}(R,z) = \phi_R(R,0) + \phi_z(R,z),\\
		\phi_z(R,z) \eee 4\pi G\int_0^z dz'\int_0^{z'} dz'' \rhod(R, z'') ,\\
		\phi_R(R,0) = -\pi G \Sigma_0 R\big[ I_0\big(\frac{R}{2R_0}\big)K_1\big(\frac{R}{2R_0}\big)-I_1\big(\frac{R}{2R_0}\big)K_0\big(\frac{R}{2R_0}\big)\big],
	\end{cases}
\end{eqnarray}
where the expression for $\phi_R(R,0)$ is exact for an exponential surface density profile \eref{eqn:Expo_Surface_Den}. Here $I_i$ and $K_i$ are the modified Bessel functions of the first and second kinds respectively. 

The deprojection of an axisymmetric disc surface density $\Sigma(R)$ requires solving for the scale height $h(R)$. In the presence of non-disc background components, the $z$-dependence of $\rhod(R,z)$ is determined by the appropriate Poisson equation and vertical hydrostatic equilibrium condition \citep[e.g.][]{Spitzer1942ApJ, Kruit1981A&A}
\begin{eqnarray}\label{eqn:Hydrostatic_Poisson}
	\begin{cases}
		\frac{\pd^2(\phi_\text{d}+\phi_\text{bg})}{\pd z^2} =  4\pi G (\rhod+\rhobgeff) ,\\
		-\frac{\sigma_z^2}{\rhod}\scalebox{0.8}{\Big(}\frac{\pd \rhod}{\pd z}\scalebox{0.8}{\Big)} = \frac{\pd (\phi_\text{d}+\phi_\text{bg})}{\pd z},
	\end{cases}
\end{eqnarray}
where $\phi_\text{bg}$ denotes the background gravitational potential. The \textit{effective} background density \citep{Lacey1985ApJ}
\begin{eqnarray}\label{eqn:rho_bg_eff}
	\rhobgeff(R) \eee  \rho_\text{bg}(R) - \frac{\pd \vc^2/\pd R}{4\pi G R},
\end{eqnarray}
accounts for contributions from both the local \textit{non-disc} background density $\rho_\text{bg}$ as well as the enclosed mass distribution (including non-local disc mass) expressed in terms of the total circular velocity $\vc(R)$; the latter term vanishes for a flat rotation curve. The $z$-dependence of the second term on the RHS of \eref{eqn:rho_bg_eff} is negligible for $z \ll R$, as the sub-leading corrections are of order $z^2/h^2, z^2/R^2,$ and $z^2/hR$ \citep{Bahcall1984ApJ}.

Equation (\ref{eqn:Hydrostatic_Poisson}) can be solved exactly when the disc self-gravity dominates (SGD) $\rhod\gg \rhobgeff$ or in the background-dominated (BGD) limit $\rhod \ll \rhobgeff$, yielding the following closed-form expressions 
\begin{eqnarray}\label{eqn:Disc_Den_Profiles}
	\rhod(R,z)=
	\begin{cases}
		\frac{\Sigma(R)}{2 h(R)} \text{sech}^2[z/h(R)]\;\;\;\;\;\;\;\;\;\;\:\text{(SGD)},\\
		\frac{\Sigma(R)}{2 h(R)} e^{-\pi z^2/[4 h^2(R)]}\;\;\;\;\;\;\;\;\;\:\text{(BGD)},
	\end{cases}
\end{eqnarray}
and corresponding disc scale heights
\begin{eqnarray}\label{eqn:Scale_Height_Limits}
	h(R) = 
	\begin{cases}
		\frac{\sigma_z^2(R)}{\pi G \Sigma(R)}\;\;\;\;\;\;\;\;\;\;\;\;\;\;\;\;\: \text{(SGD)},\\
		\frac{\sigma_z(R)}{\sqrt{8 G \rhobgeff(R)}} \;\;\;\;\;\;\;\;\;\;\;\:\: \text{(BGD)}.
	\end{cases}
\end{eqnarray} 
A locally isothermal sheet of disc stars becomes more concentrated around the mid-plane as the vertical gravity increases. Given a projected surface density profile $\Sigma(R)$, \eref{eqn:Scale_Height_Limits} computes the vertical scale height considering only the disc self-gravity (SGD) or non-disc background contributions (BGD). The transition between these two limits should be continuous, which can be achieved by equating the two expressions of \eref{eqn:Scale_Height_Limits} \citep{Lacey1985ApJ}  \footnote{In the intermediate regime $\sigma_z \sqrt{8 G \rhobgeff}\big/(\pi G \Sigma) \simeq 1$ where the total vertical gravity is comparably sourced by the disc and background, assuming either the SGD or BGD limit undercounts the local vertical gravity. The analytical expressions \eref{eqn:Scale_Height_Limits} overestimate the disc scale heights observed in simulations by about a factor of 2 in this regime (Yang et al., in preparation).}
\begin{eqnarray}\label{eqn:sigma_z_transition}
	\sigma_z(R)  = 
	\frac{\pi G \Sigma(R)}{\sqrt{8 G \rhobgeff(R)}};
\end{eqnarray} 
LHS < RHS corresponds to the SGD limit, and vice versa. The vertical disc potential following the decomposition in \eref{eqn:Disc_Gra_Potential} reads
\begin{eqnarray}\label{eqn:Vertical_Potentials}
	\phi_z= 
	\begin{cases}
		2\pi G \Sigma h\times \log\big[ \text{cosh}(z/h)\big]\;\;\;\;\;\;\;\;\;\;\;\;\;\;\;\;\;\;\;\;\;\;\;\:\:\:\text{(SGD)},\\		
		4G \Sigma h \big[e^{-\pi z^2/(4 h^2)}-1 + \big(\frac{\pi z}{2h}\big)  \erf\big(\frac{\sqrt{\pi}z}{2h}\big)\big] \:\:\:\:\text{(BGD)},
	\end{cases}
\end{eqnarray}
where erf denotes the error function. We will show in Sec. \ref{ssec:Heat_Master_Eq} that the distinct $z$-dependence of $\rhod$ and $\sigma_z$-dependence of $h$ in these two limits result in markedly different disc heating behaviour.

\subsection{Baryon Mass Infall and Disc Inside-out Formation}\label{ssec:Disc_Mass_Accretion}

\begin{figure}
		\includegraphics[width=0.98\linewidth]{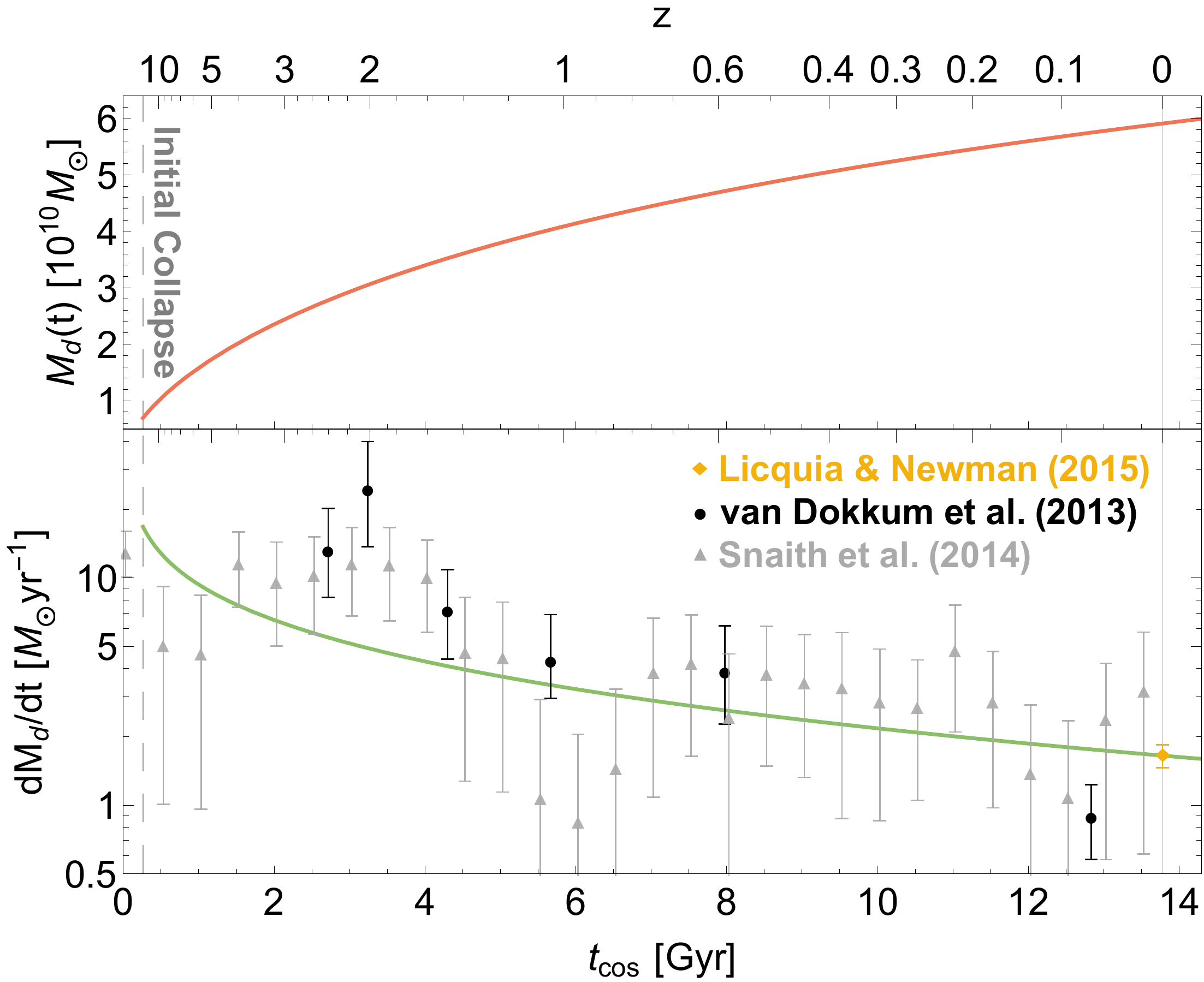}
	\caption{Total disc mass (top) and baryon infall rate (bottom), anchored to the current disc star formation rate $1.65\pm0.19$ M$_\odot$yr$^{-1}$ (\citealp{Licquia2015ApJ806}; yellow diamond). This profile is compatible with measurements by \citet{vanDokkum:2013hza} (black circles) and \citet{Snaith2014ApJ781L} (grey triangles).}
	\label{fig:MW-Disc_Mass_Infall}
\end{figure}

The secular evolution of the Galactic disc is also shaped by the continuous star formation. To estimate for the baryon mass infall, we adopt the spherical density perturbation model in a $\Lambda$CDM universe by \citet{Gunn:1972sv} and anchor the mass accretion profile to the current disc star formation rate $1.65\pm0.19$ M$_\odot$yr$^{-1}$ (\citealp{Licquia2015ApJ806}; yellow diamond). Taking $H_0 = 67.4$~km~s$^{-1}$Mpc$^{-1}, \Omega_{\Lambda} = 0.689, \Omega_\text{m} = 0.315,$ and $\Omega_\text{r} = 5.44\times 10^{-5}$ \citep{Planck:2018vyg}, we compare in Fig. \ref{fig:MW-Disc_Mass_Infall} the redshift-dependent disc total mass (top panel) and the baryon infall rate (bottom panel) compatible with independent measurements inferred from high-redshift MW-like galaxies (\citealp{vanDokkum:2013hza}; black circles) and the chemical abundance evolution of low-$[\alpha/\text{Fe}]$ stars (\citealp{Snaith2014ApJ781L}; grey triangles). The initial spherical collapse took place at $\simeq 0.3$ Gyr after the Big Bang (vertical grey dashed line) and half of the present-day disc mass was assembled by $10.7$ Gyr ago, largely consistent with the recent age measurement by \citet{Xiang2022Nature}

The early structure formation and assembly history of the Galactic disc encode information indispensable to accurately track the time-dependence of the surface density profile $\Sigma(R,t)$. \citet{vanDokkum:2013hza} establish empirically from progenitors out to high redshifts $\redz \lesssim 2.5$ that stellar discs in MW-like galaxies follow an inside-out formation pattern. The observed age and metallicity gradients of the Galactic disc stars extracted from recent high-precision survey data of \textit{Gaia}, SEGUE, APOGEE, and Kepler are also consistent with the inside-out growth history \citep[e.g.][]{Frankel2019ApJ884F,Katz2021A&A655A,Lu2022MNRAS512}. Following the simplified treatment in \citetalias{Church2019MNRAS}, we assume a constant central surface density $\Sigma_0$, and the continuous mass infall increases only the disc scale radius $R_\text{d}(t)$ till reaching the current value $R_0 = 3.2$ kpc. The accreted disc mass $M_\text{d}(t)$ and the logarithmic derivative of $\Sigma(R,t)$ read \citepalias{Church2019MNRAS}
\begin{eqnarray}\label{eqn:Disc_Mass_Growth_Profile}
	\begin{cases}
		M_\text{d}(t) = M_\text{d}(t_0) \scalebox{0.9}{\Big[}\frac{R_\text{d}(t)}{R_0}\scalebox{0.9}{\Big]}^2,\\
		\frac{d}{dt}\log\Sigma(R,t) = \frac{1}{2}\big(\frac{R}{R_0}\big)\scalebox{0.9}{\Big[}\frac{dM_\text{d}(t)/dt}{M_\text{d}(t)}\scalebox{0.9}{\Big]}\scalebox{0.9}{\Big[}\frac{M_\text{d}(t_0)}{M_\text{d}(t)}\scalebox{0.9}{\Big]}^{1/2},
	\end{cases}
\end{eqnarray}
where $M_\text{d}(t_0) = 5.9\times 10^{10}$ M$_\odot$ is the current disc total mass. The gradual increase in $\Sigma(R,t)$ also indicates that a sizeable fraction of the disc stellar populations were first born in a BGD background.

\begin{figure}
	\includegraphics[width=0.98\linewidth]{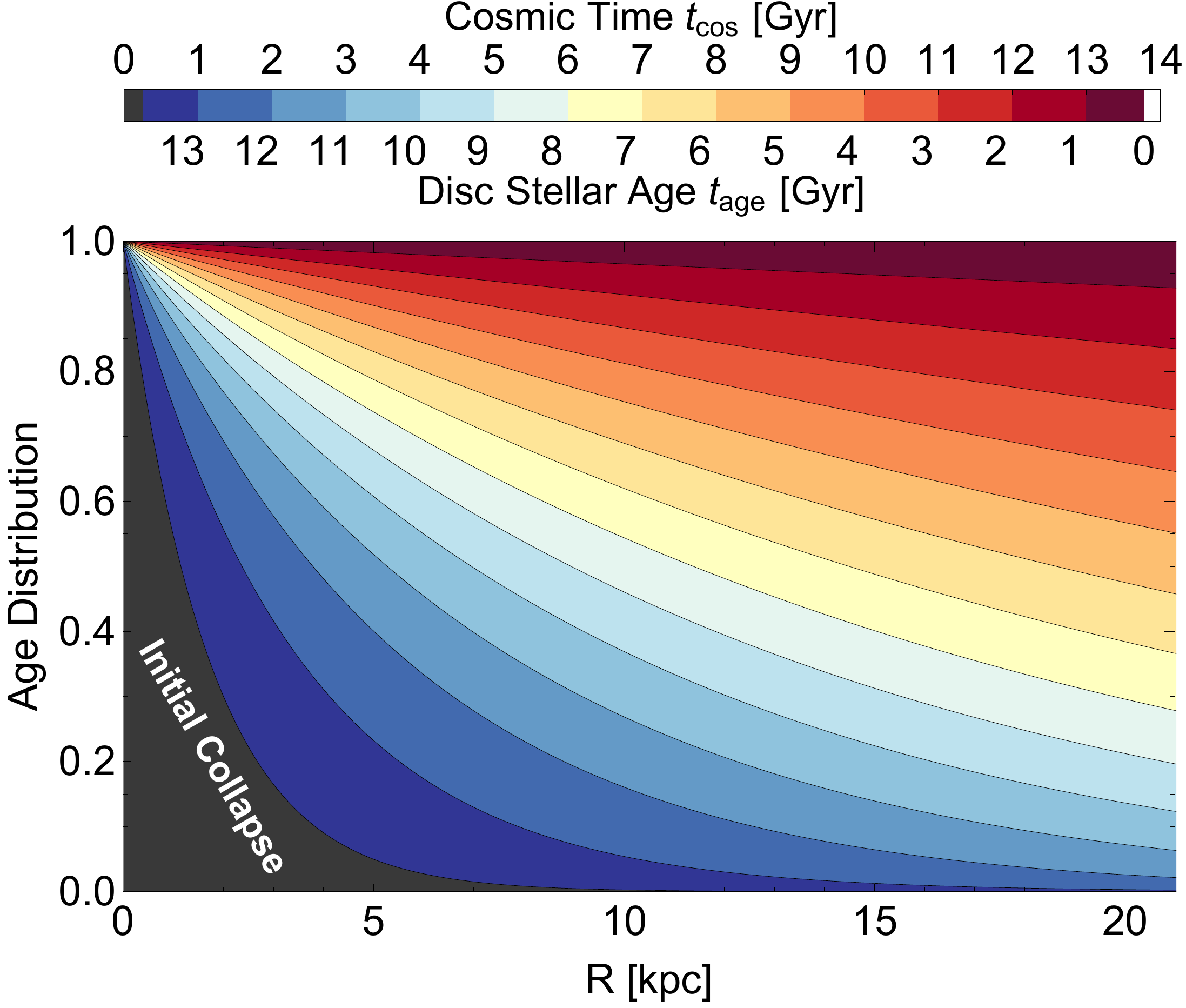}
	\caption{Age distribution of disc stellar populations predicted by the mass
		infall profile (Fig. \ref{fig:MW-Disc_Mass_Infall}) following an inside-out growth pattern \eref{eqn:Disc_Mass_Growth_Profile}.}
	\label{fig:Disc_Age_Distribution}
\end{figure}

Figure \ref{fig:Disc_Age_Distribution} shows the resulting disc stellar age distribution on the $y$-axis (normalised to unity) at various radii $R$ from the Galactic centre. This reconstructed profile correctly reflects that the spatial distribution of high-$[\alpha/\text{Fe}]$ disc stars is more radially concentrated than the low-$[\alpha/\text{Fe}]$, young populations \citep{Bovy2012ApJ753}. Stellar populations in the inner disc $R\lesssim 4$ kpc have ensemble-averaged ages $\overline{t}_\text{age} \gtrsim 10$ Gyr. The mean stellar age decreases monotonically with the Galactocentric radius, dropping to $\overline{t}_\text{age} \simeq 5.5$ Gyr at $R = 20$~kpc. The thick and thin disc components are expectedly associated with distinct dynamical heating timescales. We have not incorporated radial migration, of which the effect is further discussed in Sec. \ref{ssec:FDM_Heating_Predictions}.


\subsection{Realistic MW Background and Observational Constraints}\label{ssec:MW_Background_Disc_sigma_z}

The MW non-DM profile mainly consists of the black hole Sgr A* ($4.15\times 10^6$~M$_\odot$; \citealp{Abuter2019A&A}), a nuclear star cluster (NSC; $6.1\times 10^7$~M$_\odot$; \citealp{Gallego-Cano2020A&A}), nuclear star disc (NSD; $6.9\times 10^8$~M$_\odot$; \citealp{Gallego-Cano2020A&A}), B/P bulge ($8.9\times 10^9$~M$_\odot$; \citealp{McMillan2017MNRAS}), and the Galactic disc. To accurately reconstruct the Galactic disc height via \eref{eqn:Scale_Height_Limits}, we adopt the best-fit $\sigma_z(R,z,[\text{Fe}/\text{H}],\tage)$ profile by \citet{Sharma2021MNRAS506} derived from $\simeq 840,000$ dynamical tracers. This multi-variable function is observationally robust for $1\text{ kpc} \lesssim R \lesssim 18$ kpc and consistent with other recent measurements \citep[e.g.][]{Mackereth2019MNRAS, Hayden2020MNRAS493}. Recognising the non-trivial dependence of stellar metallicity $[\text{Fe}/\text{H}]$ and vertical height from the mid-plane $z$ on the population age $\tage$ and $R$ \citep{Feuillet2019MNRAS489, Miglio2021A&A645A}, to a first approximation we adopt the $R$-dependent mean metallicity derived in \citet{Hayden2014AJ147}. The stellar age distribution at each radius $R$ is inferred from the mass infall model introduced in Sec.~\ref{ssec:Disc_Mass_Accretion} (see Fig.~\ref{fig:Disc_Age_Distribution}). We further associate the thick disc populations (with age $\geq 8$~Gyr) with the scale height $z = 0.7$~kpc and thin disc stars ($\tage\leq 8$~Gyr) with $z = 0.3$ kpc \citep{Bovy2012ApJ753}. This age-distribution-weighted dispersion profile yields $\sigma_z(R_\odot) \simeq 22$~km~s$^{-1}$. Extrapolation beyond $18$ kpc becomes ill-constrained due to the limited number of dynamical tracers and decline in apparent brightness. Disentangling the disc and bulge populations within the central 1~kpc is intrinsically challenging. We hence exclude $R \lesssim 1$~kpc from our analysis, where the disc morphology and kinematic information are ill-constrained.

\begin{figure}
	\centering
	\includegraphics[width=0.99\linewidth]{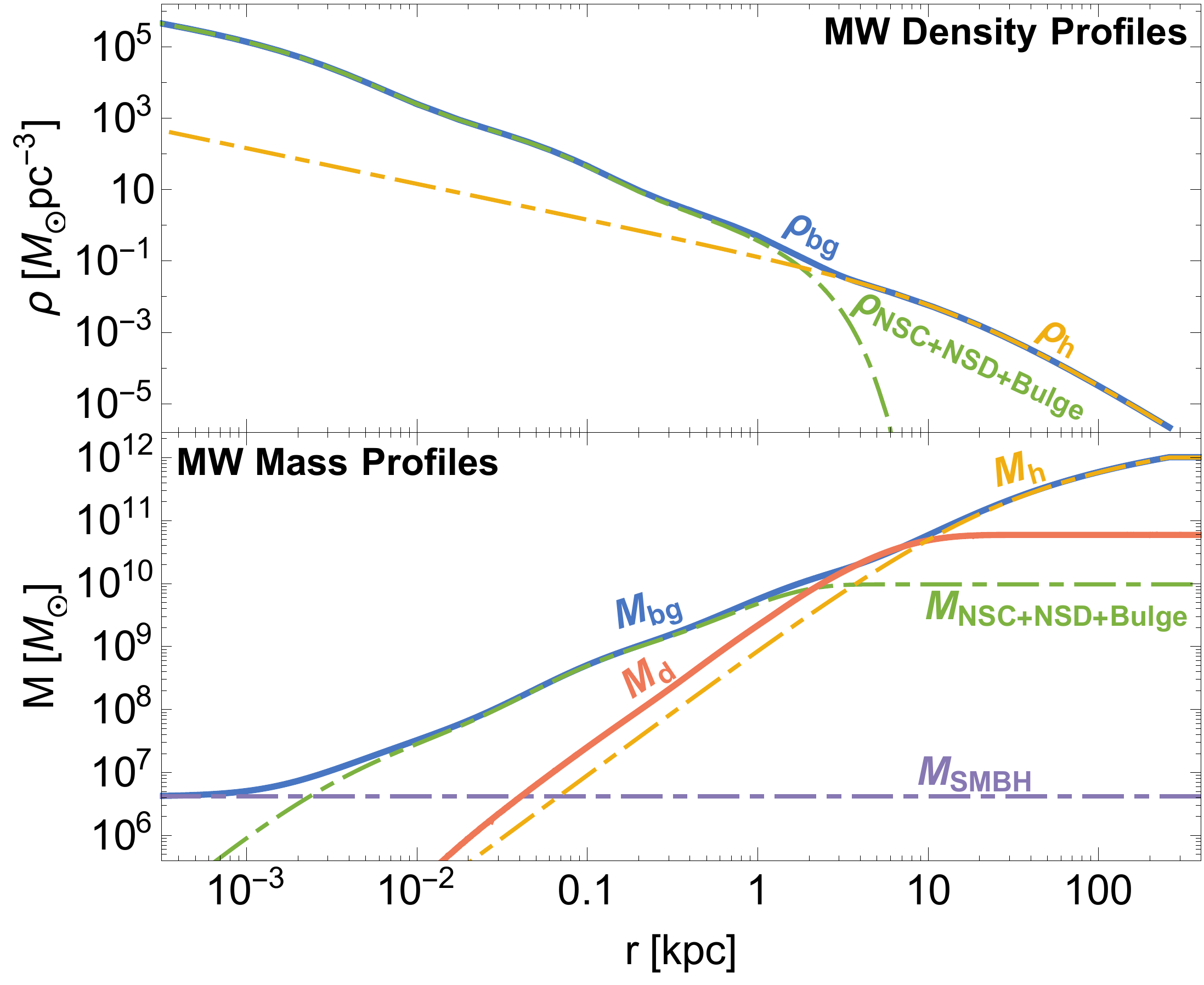}
	\caption{Shell-averaged density (top) and enclosed mass profiles (bottom) of the MW DM halo (yellow), central baryon structures (green; including the NSC, NSD, and B/P Bulge), and Sgr A* (purple). The MW background (blue) accounts for all the non-disc components. The Galactic disc total mass enclosed within the axial radius $R$ is shown by the solid red curve.}
	\label{fig:MW-Total}
\end{figure}

Current inferences of the MW DM halo virial mass still exhibit an appreciable spread $0.5$\textendash$2\times 10^{12}$~M$_\odot$ \citep[e.g.][and references therein]{Callingham:2019vcf, Wang:2019ubx}. Here we instead adopt the shell-averaged density and velocity dispersion profiles of a self-consistently constructed isotropic MW-sized FDM halo; the relaxed halo is well fitted by an NFW profile with $\Mh = 1.0\times 10^{12}$~M$_\odot,$ concentration parameter $c = 15,$ virial radius $r_\text{vir} = 260$~kpc, and a 1D velocity dispersion $\sigmah(R_\odot) = 86$~km~s$^{-1}$ in the solar neighbourhood (Su et al., in preparation; \citealp{Lin:2018whl}). For the range of $m_a = 0.3$\textendash$5.0\times 10^{-22}$~eV considered in this work, possible soliton core sizes $\rc$ are sub-kpc, indicating soliton-induced dynamical effect in the MW dominates only within $r \lesssim 1$~kpc. We therefore do not include the modelling of $m_a$-dependent FDM soliton in the non-disc background. (see Sec.~\ref{sssec:Core-halo_Soliton_Size} for a detailed discussion). Figure~\ref{fig:MW-Total} shows the shell-averaged density (top) and mass profiles (bottom).

\begin{figure}
	\begin{center}
		\includegraphics[width=0.98\linewidth]{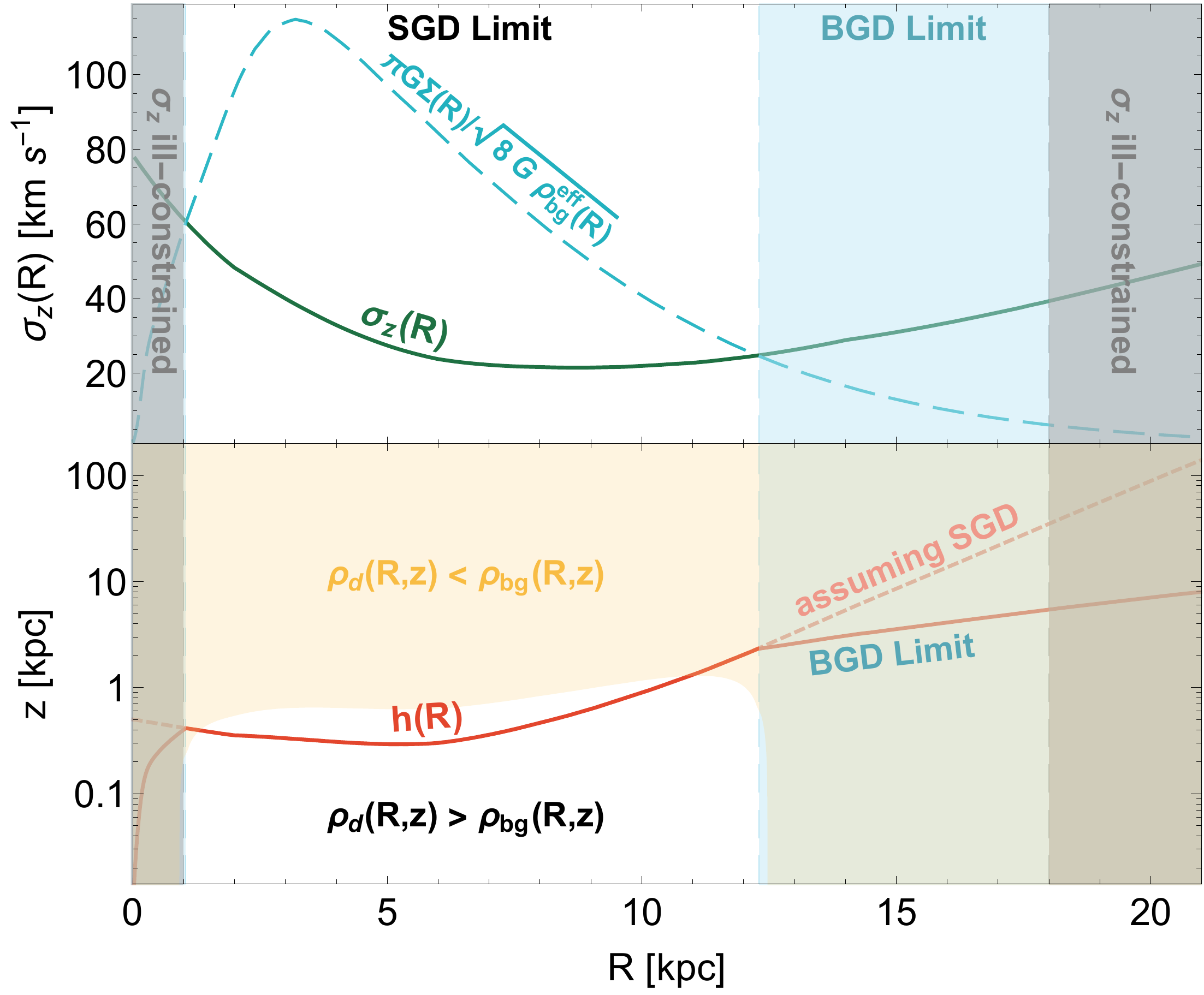}
	\end{center}
	\caption{\textit{Top:} Comparing the values of age-distribution-weighted disc vertical velocity dispersion by \citet{Sharma2021MNRAS506} (green; ill-constrained in the grey-shaded areas including $R \lesssim 1$~kpc and $R\gtrsim 18$~kpc; see Sec. \ref{ssec:MW_Background_Disc_sigma_z}) with the RHS of \eref{eqn:sigma_z_transition} (cyan) at each radius $R$ determines whether locally the disc self-gravity-dominated (SGD; transparent) or background-dominated (BGD; cyan shaded) limit is applicable. \textit{Bottom:} Galactic disc scale height (solid red) given by \eref{eqn:Scale_Height_Limits} that accounts for the disc self-gravity and background mass distribution. In the BGD limit, consistent with regions where the genuine total background density $\rho_\text{bg}$ exceeds the disc density $\rhod$ (yellow shaded), neglecting non-disc mass components would severely overestimate the scale height (dashed light red). The present Galactic disc is self-gravitating only for $1$ kpc $\lesssim R\lesssim 12$~kpc.}
	\label{fig:MW-Disc_sigma_z-h(R)}
\end{figure}

Figure \ref{fig:MW-Disc_sigma_z-h(R)} in the top panel compares the resulting $\sigma_z$ (green) with the RHS of \eref{eqn:sigma_z_transition} (cyan); the two curves intersect at the transition between the disc self-gravity-dominated (SGD; transparent) and background-dominated (BGD; cyan shaded) regimes. The striking `U-shaped' $\sigma_z(R)$ profile indicates non-uniform stellar heating rates across the entire disc, as the ensemble-averaged stellar ages decrease appreciably with $R$ (Fig. \ref{fig:Disc_Age_Distribution}). Near the solar neighbourhood $\sigma_z(R)$ reaches the global minimum. The formation of inner thick disc and the outer disc flaring of young, low-$[\alpha/\text{Fe}]$ populations thus originates from distinct sources of stellar heating. This kinematic signature can be naturally produced by gravitational interactions between disc stars and FDM substructures, as addressed in Sec. \ref{sec:Dynamical_Evolution_Disc}.

The bottom panel of Fig. \ref{fig:MW-Disc_sigma_z-h(R)} shows the disc scale height (red) appropriate in either the SGD or BGD limit, giving $h(R_\odot) = 0.49$~kpc in the solar vicinity. The disc morphology is dictated by the background gravitational potential when $\rhod(R,z) < \rho_\text{bg}(R,z)$ (yellow shaded), consistent with the SGD/BGD limit determined by \eref{eqn:sigma_z_transition}. The stellar disc dominantly sources the total gravitational potential only in the region $1$~kpc $\lesssim R \lesssim 12$ kpc. Accounting for the inner baryon mass distribution (outer DM halo) for $R\lesssim1$ kpc ($\gtrsim 12$ kpc) is thus pivotal in properly determining the disc vertical structure. We emphasise that the outer profile of disc scale height $h(R\geq 18$~kpc) is likely unreliable, where $\sigma_z$ becomes ill-constrained. For the same reason $R\leq 1$ kpc is excluded in our analysis.

\section{Dynamical Evolution of the Galactic Disc}\label{sec:Dynamical_Evolution_Disc}
	
	Gravitational encounters with local DM substructures could leave observable imprints on the Galactic disc stars over a Hubble time. In Sec. \ref{ssec:Heat_Master_Eq}, the adiabatic invariance of disc stars' vertical action is exploited to derive the master equation for stellar heating suitable in either the SGD or BGD limit. We quantify the disc heating rates induced by FDM halo density granulation in Sec. \ref{ssec:FDM_Halo_Granulation}, and passages of FDM subhaloes in Sec. \ref{ssec:FDM_Subhalo_Passages}. In a parallel CDM simulation of disc heating as discussed in Sec.~\ref{ssec:CDM_Counterparts}, the DM subhalo component would be stronger but the DM granularity of FDM does not exist. Thus the heating effects in the FDM and CDM paradigms are quite distinct.

\subsection{Master Equation for Disc Heating}\label{ssec:Heat_Master_Eq}

Consider a background of self-similar perturbers of mass $\Mp(r) \gg \bigO(1)$ M$_\odot$, velocity $\vp(r)$, number density $n(r)$, and a typical size $b_\text{min}(r)$, embedded in a spherically symmetric DM halo. The resulting dynamical effect on a test star with velocity $\vbf$ can be approximately reformulated as a classical two-body relaxation process, valid in the \textit{diffusive} regime where the stellar dynamical length scale is large compared to $b_\text{min}$ \citep[e.g.][]{Binney&Tremaine2008}. For DM overdensities and stellar populations exhibiting spatially uncorrelated velocities, a perturber passage with an impact parameter $b$ causes the direction-averaged trajectory deflection of a disc star $\<\vbf\cdot \Delta \vbf\> \simeq 0$. The mean square increase per weak gravitational encounter $\<\Delta v^2\> \simeq (G\Mp/b\vp)^2$ results in a stellar heating rate (time rate of change of energy per unit mass) $\simeq \int_{b_\text{min}}^{b_\text{max}} db(2\pi b) n \vp\<\Delta v^2\>$. For a disc star with a vertical oscillation period $P(R, \zmax)$ determined by the maximum vertical displacement $\zmax$, adiabatic transits $b/\vp > P/2$ incur no long-term energy exchange. Adiabatic cutoff $b_\text{max}= P\vp/2$ \footnote{\label{fn:Coulomb_Logarithm}For stars or dispersion-supported systems characterised by a single dynamical timescale, the maximum impact parameter can be suitably taken as the orbital radius $b_\text{max} = r$  \citep{Bar-Or2019ApJ, Chowdhury2021ApJ}. In contrast, for a disc star the distinct timescales associated with nearly circular ($\hat{\phi}$) motion, epicyclic ($\hat{r}$), and vertical ($\hat{z}$) oscillations yield three different Coulomb logarithms corresponding to the respective adiabatic cutoff scales.} then demarcates the relevant domain of impact parameter $b\in [b_\text{min}, b_\text{max}]$. The locally homogeneous background of identical perturbers gives a \mbox{vertical heating rate \citep{Lacey1985ApJ}}
\begin{eqnarray}\label{eqn:Transient_Perturbation_Rate}
		\bigH = \frac{G^2\Mp^2}{\vp} 4\sqrt{2}\pi n \log\Lambda \eee \bigM \log\bigg(\frac{P}{\tau}\bigg),
\end{eqnarray}
where $\log\Lambda \eee \log(b_\text{max}/b_\text{min})$ is the Coulomb logarithm. Here $\bigM\eee 4\sqrt{2}\pi n(G^2 \Mp^2/\vp)$ and $\tau \eee 2 b_\text{min}/\vp$.

The limit of Coulomb logarithm approaching zero $\ln\Lambda \rightarrow 0$ corresponds to the strongly coupled regime in plasma physics. Strong-collision-dominated interactions render many linearisation techniques (e.g. the Fokker\textendash Planck approximation in deriving leading-order diffusion coefficients) invalid. In an FDM halo, the vertical heating of disc stars now vanishes due to adiabatic invariance $P\lesssim\tau$. Furthermore, a near-zero or negative Coulomb logarithm $\log(r/b_\text{min})$ indicates that a test star or dispersion-supported system no longer travels across spatially uncorrelated patches of potential fluctuations per orbit. Since the self-interaction of background perturbers is no longer negligible, halo granules cannot be approximated as travelling on straight trajectories at a constant velocity. Assessing heating effectiveness in this \textit{non-diffusive} regime as the test object is embedded in a local DM substructure requires self-consistent simulations. We exclude possible non-diffusive energy gain by replacing $\ln\Lambda \leq 0$ with $\ln\Lambda = 0$ as a conservative estimate.

Individual member stars of a locally isothermal ensemble at each radius $R$ can be characterised by the respective maximum vertical displacement from the mid-plane $\zmax$ as determined by \eref{app:Jz_P}. The vertical oscillation period of a disc star reaching $\zmax$,
\begin{eqnarray}\label{eqn:P_Definition}
	P(R,\zmax) = 2\int_{-\zmax}^{\zmax} \frac{dz}{\sqrt{2[\phi_\text{tot}(R,\zmax)-\phi_\text{tot}(R,z)]}},
\end{eqnarray}
depends on both the disc and background gravitational potentials $\phi_\text{tot} \eee \phi_\text{d}+\phi_\text{bg}$. We have numerically verified that the MW non-disc background yields $\phi_\text{bg}$ harmonic in $z$ to a good approximation, for $R\geq 1$~kpc and $z \leq 5$~kpc. The oscillation periods in either the SGD or BGD limit read
\begin{eqnarray}\label{eqn:Disc_Osc_Period}
	P(R,\zmax) =
	\begin{cases}
		\frac{\sigma_z}{\pi G \Sigma} I_{P1}(\zmax/h)\;\;\;\;\;\;\;\;\;\;\;\;\;\;\text{(SGD)},\\
		2I_{P2}(R,\zmax) \;\;\;\;\;\;\;\;\;\;\;\;\;\;\;\;\;\:\:\:\text{(BGD)},
	\end{cases}
\end{eqnarray}
where $I_{P1}(x) \eee \int_{-x}^{x} \frac{dx'}{\sqrt{\log(\cosh x/\cosh x')}}$ is a strictly monotonically increasing function bounded below by $\sqrt{2}\pi$, and $I_{P2}(R,\zmax)\eee \int_{-\zmax}^{\zmax} \frac{dz}{\sqrt{2[\phi_\text{bg}(R,\zmax)-\phi_\text{bg}(R,z)]}}$ has to be evaluated numerically. The disc scale height $h(R)$ is given by \eref{eqn:Scale_Height_Limits}.

For a stellar disc kinematically decoupled in the vertical and radial directions, the ensemble-averaged time rates of change in the disc vertical velocity dispersion squared in SGD and BGD limits are
\begin{eqnarray}\label{eqn:Heating_Eq_SGD_BGD_Limits}
	\frac{d\sigma_z^2}{dt} = 
	\begin{cases}
		\frac{2\sigma_z^2}{3}\frac{d\log\Sigma}{dt}+\kappa_1 \<\bigH\>_1 \;\;\;\;\;\;\;\;\;\text{(SGD)},\\
		\kappa_2\<\bigH\>_2 \;\;\;\;\;\;\;\;\;\;\;\;\;\;\;\;\;\;\;\;\;\;\;\;\;\;\:\:\:\text{(BGD)},
	\end{cases}
\end{eqnarray}
where the two dimensionless coefficients are $\kappa_1 = 0.526$ and $\kappa_2 = 1$. For $i = 1, 2$ we have
\begin{eqnarray}\label{eqn:Heating_Eq_SGD_BGD_Limits_2}
	\begin{cases}
		\<\bigH\>_i  \eee \bigM\log\Big(\frac{\<P\>_i}{\tau}\Big),\\
		\<P\>_1 = \frac{\sigma_z I_{\Lambda1}}{\pi G \Sigma},\\
		\<P\>_2 = 2I_{P2},
	\end{cases}
\end{eqnarray}
where $I_{\Lambda1} = 4.82$ and $I_{P2}$ is evaluated numerically with \eref{eqn:Disc_Osc_Period}. See Appendix \ref{app:disc_heating_derivation} for detailed derivations. The equality $\<P\>_2 = 2I_{P2}$ assumes a $\zmax$-independent $I_{P2}$, which is strictly valid only for harmonic vertical potential $\phi_\text{bg} \propto z^2$. In the MW non-disc background, the vertical oscillation periods of disc stars reaching $0.5$~kpc~$\leq \zmax \leq 5$~kpc differ by less than $\simeq 3\%$ ($0.8\%$) at $R_\odot$ ($R = 20$~kpc), validating the approximation adopted here.

In the SGD and BGD limits, \eref{eqn:Heating_Eq_SGD_BGD_Limits} results in two distinct pictures of the stellar disc heating. When the background potential is comparatively negligible, $\kappa_1 \simeq 0.5$ can be interpreted as a manifestation of the virial theorem \citepalias{Church2019MNRAS}. Continuous mass accretion also contributes to disc heating via the term $d\log\Sigma/dt$, as the newly formed stellar populations deposit the potential energy acquired during infall into the self-bound disc. The logarithmic derivative of Galactic disc surface density is given by \eref{eqn:Disc_Mass_Growth_Profile}. On the other hand, the BGD limit receives a relative enhancement in the gravitational heating efficiency $\kappa_2 = 1$. To remedy the unphysical discontinuity in the disc heating behaviour at the SGD and BGD limit transition, we directly apply \eref{eqn:Heating_Eq_SGD_BGD_Limits} only when $\frac{\sigma_z \sqrt{8 G \rhobgeff}}{\pi G \Sigma} \leq 0.5$ ($\geq 1.5$) in the SGD (BGD) limit, corresponding respectively to $\frac{\phi_\text{d}(R,\<\Delta z\>)}{\phi_\text{bg}(R,\<\Delta z\>)} \gtrsim 5$ ($\lesssim 0.2$); see \eref{eqn:sigma_z_transition}. In the intermediate regime $0.5 < \frac{\sigma_z \sqrt{8 G \rhobgeff}}{\pi G \Sigma} < 1.5$, we compute and linearly interpolate the heating rates in both limits \eref{eqn:Heating_Eq_SGD_BGD_Limits}. For instance, the resulting heating rate for $\frac{\sigma_z \sqrt{8 G \rhobgeff}}{\pi G \Sigma} = 1.0$ is the arithmetic mean of those in the SGD and BGD limits.

For a sufficient large Coulomb factor $\<P\>_i/\tau \gg 1$, the stellar heating rate $\<\bigH\>_i$ is comparatively insensitive to $\<P\>_i$ and roughly scales with $\bigM$ for the physical process of interest. The accumulative energy transfer on a disc star sourced by repeated weak gravitational encounters with FDM halo granules and orbiting subhaloes is examined in Secs. \ref{ssec:FDM_Halo_Granulation} and \ref{ssec:FDM_Subhalo_Passages}, with the respective $\bigM$ defined in \pCAeref{eqn:Granulation_Heating_Rate}\pCCeref{eqn:Subhalo_Heating_Rate}.
The possibly non-trivial disc heating due to satellite accretion \citep[e.g.][]{Toth1992ApJ389, Quinn1993ApJ403} is not accounted for in the present calculations \pCAeref{eqn:Heating_Eq_SGD_BGD_Limits}\pCCeref{eqn:Heating_Eq_SGD_BGD_Limits_2}.

\subsection{FDM Halo Granulation}\label{ssec:FDM_Halo_Granulation}
Quantum wave interference inside an FDM halo gives rise to ubiquitous, stochastically fluctuating substructures of sizes comparable to the de Broglie wavelength $\lambda_\text{dB}$ \citep[e.g.][]{Schive:2014dra, Veltmaat:2018dfz}. The resulting dynamical effect on stellar populations can be captured analytically by the classical kinetic theory \citep{Bar-Or2019ApJ, El-Zant:2019ios, Chavanis2021EPJP136}, a framework that is recently verified with FDM simulations by \citet{Chowdhury2021ApJ} in the regime where diffusive heating dominates over or is balanced by dynamical friction.

The first-order and second-order diffusion coefficients induced by gravitational encounters between a test star $m$ and the stochastic density fluctuations inside an FDM halo with a Gaussian-distributed velocity dispersion $\sigmah$ read \citep{Bar-Or2019ApJ} \footnote{\label{fn:Halo_Velocity_Dispersion} The \textit{genuine} halo 1D velocity dispersion $\sigmah$ here is equivalent to $\sigma_\text{eff}$ defined in \citet{Bar-Or2019ApJ}. The quantity $\sigma \eee \sqrt{2}\sigmaeff = \sqrt{2}\sigmah$ in \citet{Bar-Or2019ApJ} factually corresponds to the \textit{effective} velocity dispersion obtained from solving the Jeans equation; see \citet{Chowdhury2021ApJ} for a detailed discussion.}
\begin{eqnarray}\label{eqn:FDM_Diffusion_Coeff_1}
	\begin{cases}
		vD[\Delta v_\parallel] = -\bigD\Xeff\scalebox{0.9}{\Big[}\bigG(\Xeff)+\frac{m}{2\Mgra}\bigG\scalebox{0.9}{\Big(}\frac{\Xeff}{\sqrt{2}}\scalebox{0.9}{\Big)}\scalebox{0.9}{\Big]},\\
		D[(\Delta v_\parallel)^2] = \bigD\frac{\bigG(\Xeff)}{\Xeff},\\
		D[(\Delta \vbf_\bot)^2] = \bigD\frac{\erf(\Xeff)-\bigG(\Xeff)}{\Xeff},
	\end{cases}
\end{eqnarray}
where $\Xeff \eee \frac{v}{\sqrt{2}\sigmah}$ and
\begin{eqnarray}\label{eqn:FDM_Diffusion_Coeff_2}
	\begin{cases}
		\bigD= \frac{4\sqrt{2}\pi G^2 \rhoh \Mgra }{\sigmah}\log\Lambda,\\
		\bigG(x) \eee \frac{1}{2x^2}\big[\erf(x)-\frac{2x}{\sqrt{\pi}}e^{-x^2}\big].
	\end{cases}
\end{eqnarray}
The typical effective mass of FDM density granulation $\Mgra(r)$ can be estimated analytically \citep{Bar-Or2019ApJ}
\begin{eqnarray}\label{eqn:M_gra}
	\begin{cases}
		\Mgra \eee \rhoh\scalebox{0.9}{\Big(}\frac{\lambda_\text{dB}}{2\sqrt{\pi}}\scalebox{0.90}{\Big)}^3,\\
		\lambda_\text{dB} = \frac{2\pi\hbar}{m_a(\sqrt{2}\sigmah)},\\
		\Lambda=\frac{b_\text{max}}{b_\text{min}} = \frac{P\sigmah/2}{(\lambda_\text{dB}/2\pi)/2} \eee \frac{P}{\tau},
	\end{cases}
\end{eqnarray}
where $\hbar$ denotes the reduced Planck constant. For $m_a \simeq 10^{-22}$ eV, $M_\text{gra}(R_\odot) \simeq 10^6$ M$_\odot$ in the solar neighbourhood, comparable to the mass scale of Galactic giant molecular clouds \citep[e.g.][]{Lacey1984MNRAS208, Hanninen2002MNRAS337}. The present-day ensemble-averaged vertical oscillation period $\<P(R_\odot)\> \simeq 110$ Myr and the nearby granulation dynamical timescale $\tau(R_\odot) \simeq 2$ Myr yield $\ln \Lambda \simeq 4$. The Coulomb logarithm increases mildly with $R$ due to decreasing $\Sigma(R)$ and $\rhoh(R)$.

The 1D velocity of disc stars can be reparametrised as \footnote{Strictly speaking, the azimuthal velocity squared also includes a contribution from the azimuthal dispersion $v_\phi^2 = \vc^2 + \sigma_\phi^2$. For the Galactic disc however, $\sigma_\phi \simeq 30$ km s$^{-1}$ \citep{Guiglion2015AA583A}; this addition contribution is suppressed by $(\sigma_\phi/\vc)^2 = \bigO(10^{-4})$, giving $v_\phi \simeq \vc$.}
\begin{eqnarray}\label{eqn:Definition_nu}
	\sqrt{\frac{\vc^2 + \sigma_z^2+\sigma_R^2}{3}} \simeq \sqrt{\frac{\vc^2 +3.4 \sigma_z^2}{3}} \eee \mu(R)\sigma_z(R),
\end{eqnarray}
given the high-$[\alpha/\text{Fe}]$ populations exhibit a nearly age- and $z$-independent $\frac{\sigma_z}{\sigma_R} = 0.64\pm 0.04$ \citep{Mackereth2019MNRAS}; therefore $\Xeff = \frac{v}{\sqrt{2}\sigmah} = \frac{\sqrt{3}\mu\sigma_z}{\sqrt{2}\sigmah}$. The specific energy diffusion coefficient $D[\Delta E_z]  = \frac{\sigma_z^2}{v^2}vD[\Delta v_\parallel] + \frac{\sigma_z^2}{2 v^2}D[(\Delta v_\parallel)^2]+\frac{1}{4}\Big(1-\frac{\sigma_z^2}{v^2}\Big)D[(\Delta \vbf_\bot)^2]$ in the $z$-direction \citep{Binney&Tremaine2008} can be simplified as
\begin{eqnarray}
	\begin{cases}
		D[\Delta E_z] = \bigD\frac{\bigE(\Xeff)}{4\Xeff},\\
		\bigE(\Xeff) \eee  \frac{4\Xeff}{3\mu^2\sqrt{\pi}}e^{-\Xeff^2}+(1-\mu^{-2})\big[\erf(\Xeff)-\bigG(\Xeff)\big],
	\end{cases}
\end{eqnarray}
provided that $m=\bigO(1)$~M$_\odot\ll \Mgra$ for the disc stars. In the limit $\sigma_z \ll \vc$, the expression reduces to $\bigE(\Xeff) \simeq \erf(\Xeff)-\bigG(\Xeff)$. The flow of total specific energy into disc stars' vertical motion due to FDM wave interference is then (cf. \eref{eqn:Transient_Perturbation_Rate})
\begin{eqnarray}\label{eqn:Granulation_Heating_Rate}
	\begin{cases}
		\bigH = D[\Delta E_z] = \bigM\log\big(\frac{P}{\tau}\big),\\
		\bigM \eee \frac{\pi^{5/2} G^2 \rhoh^2\hbar^3}{2m_a^3\sigmah^4} \frac{\bigE(\Xeff)}{\Xeff},\\
		\tau \eee \frac{\hbar}{\sqrt{2} m_a\sigmah^2}.
	\end{cases}
\end{eqnarray}

It is instructive to introduce the characteristic heating timescale driven by FDM halo density granulation \citep{Bar-Or2019ApJ}
\begin{eqnarray}\label{eqn:Heating_Time_Scale}
	T_\text{heat} \eee \frac{\sigmah^2}{d\sigma_z^2/dt} \propto m_a^3\sigmah^6\rhoh^{-2}
\end{eqnarray}
defined with respective to the velocity dispersion of a background FDM halo. The granule-driven disc heating rate $\bigH \propto T_\text{heat}^{-1}$ is sensitive to the FDM halo physical attributes; in particular an accurate inference of $\sigmah(R)$ is pivotal in estimating the stellar heating rate. In contrast, the $\sigma_z$-dependence enters $\bigH$ only via the vertical oscillation period in the SGD limit and is logarithmically suppressed. The heating rate is hence roughly independent of $\sigma_z(t)$.

\subsection{FDM Subhalo Passages}\label{ssec:FDM_Subhalo_Passages}

The formation of a primary halo is accompanied by the birth and additional accretion of subhaloes. Over time, the unevolved (prior to accretion) subhalo mass function (SHMF) $\frac{d\nsub}{d\log \msub}$ is altered by dynamical processes such as tidal stripping and dynamical friction; $\nsub$ and $\msub$ denote the subhalo number density and mass, respectively. The unevolved CDM SHMF is commonly computed with the extended Press\textendash Schechter  formalism \citep{Press1974ApJ187, Sheth:1999mn}; the robustness and detailed calibration thereof have been well studied using $N$-body simulations \citep[e.g.][]{Tinker:2008ff,Despali:2015yla}. Qualitatively, the formation of low-mass FDM subhaloes is suppressed below the quantum Jeans length scale $\propto m_a^{-1/2}$ \citep{Marsh:2015xka}, so the subhalo population is undoubtedly reduced in FDM than in CDM. 

We adopt a simulation-guided approach to determining FDM SHMF, and the suppression thereof relative to CDM SHMF can be parameterised as \citep{Schive:2015kza,May:2021wwp}
 \begin{eqnarray}\label{eqn:FDM_SHMF}
 	\frac{d \nsub}{d \msub}\bigg|\underset{\text{FDM}\;\;\;\;\;\;\;\;}{(\msub,\redz)} = \bigg[1+\Big(\frac{\msub}{M_0}\Big)^\alphasub\bigg]^\betasub \frac{d \nsub}{d \msub}\bigg|\underset{\text{CDM}\;\;\;\;\;\;\;\;}{(\msub,\redz)}.
 \end{eqnarray}
\citet{Schive:2015kza} find $\alphasub = -1.1, \betasub = -2.2,$ and a suppression mass scale
\begin{eqnarray}\label{eqn:M0_FDM_SHMF_Schive}
	M_0(m_a) = 1.6\times 10^{10} \bigg(\frac{m_a}{10^{-22}\text{ eV}}\bigg)^{-4/3}\text{ M}_\odot,
\end{eqnarray} 
from collisionless $N$-body simulations with FDM initial power spectra. This estimate appears compatible with the recent self-consistent FDM cosmological simulations carried out by \citet{May2022arXiv220914886M}; some pertinent caveats and reported discrepancies are elaborated in Sec.~\ref{sssec:SHMF_Tidal}. The ratio between the total CDM subhalo mass and the host halo mass $\Mh = 1.0\times 10^{12}$ M$_\odot$ at redshift zero is fixed to 0.1, inferred from the CDM subhalo population statistics in $N$-body simulations by \citet{Gao2011MNRAS410}; i.e. the following normalisation condition for the unevolved CDM SHMF is imposed at $\redz = 0$
 \begin{eqnarray}\label{eqn:CDM_SHMF_Normalisation}
 	\int_0^{\Mh}\int_0^{R_\text{vir}} d\msub dr4\pi r^2\frac{d \nsub}{d\log \msub}\bigg|_\text{CDM}\eee 0.1 \Mh.
 \end{eqnarray}
The FDM SHMF is anchored to the CDM SHMF via \eref{eqn:FDM_SHMF}. This ratio $0.1$ thus serves as an upper limit in an FDM cosmology.

A newly formed FDM (sub)halo comprises a central soliton embedded inside an NFW profile \citep{Schive:2014dra}. We assume the subhalo population follows the empirical core-halo relation \citep{Schive:2014hza} \footnote{This empirical correlation can be interpreted as the isothermality of FDM haloes \citep{Hui:2016ltb,Bar:2018acw, Bar:2019bqz}. However non-trivial statistical scatter has also been observed in self-consistent cosmological simulations \citep{Schwabe:2016rze,Chan:2021bja}; the associated modelling uncertainties are discussed in Sec. \ref{sssec:Core-halo_Soliton_Size}. Notice that \citet{Schive:2014hza} define `core mass' $M_\text{c}$ as the soliton mass enclosed within $\rc$, giving $M_\text{c} \simeq M_\text{sol}/4$.} with a universal profile transition radius $\simeq 3\rc$ \citep{Mocz:2017wlg, Chiang:2021uvt, Chowdhury2021ApJ} such that the redshift-dependent soliton half-density radius $\rc(m_a,\msub,\redz)$, total mass $M_\text{sol}$, and minimum (sub)halo mass $M_\text{sub,min}$ are
\begin{eqnarray}\label{eqn:Core_Halo_Relation}
	\begin{cases}
		\rc = \frac{1.6}{\sqrt{1+\redz}}\scalebox{0.9}{\Big[}\frac{\zeta(\redz)}{\zeta(0)}\scalebox{0.9}{\Big]}^{-1/6}\scalebox{0.9}{\Big(}\frac{\msub}{10^{9}\text{ M}_\odot}\scalebox{0.9}{\Big)}^{-1/3}\scalebox{0.9}{\Big(}\frac{m_a}{10^{-22}\text{ eV}}\scalebox{0.9}{\Big)}^{-1}\text{ kpc},\\
		\frac{M_\text{sol}}{1.3\times 10^8\text{ M}_\odot} = \sqrt{1+\redz}\scalebox{0.9}{\Big[}\frac{\zeta(\redz)}{\zeta(0)}\scalebox{0.9}{\Big]}^{1/6}\scalebox{0.9}{\Big(}\frac{\msub}{10^9\text{ M}_\odot}\scalebox{0.9}{\Big)}^{1/3}\scalebox{0.9}{\Big(}\frac{m_a}{10^{-22}\text{ eV}}\scalebox{0.9}{\Big)}^{-1},\\
		M_\text{sub,min} =  4.4\times 10^7(1+\redz)^{3/4}\Big[\frac{\zeta(\redz)}{\zeta(0)}\Big]^{1/4}\Big(\frac{m_a}{10^{-22}\text{ eV}}\Big)^{-3/2}\text{ M}_\odot,
	\end{cases}
\end{eqnarray}
where $\zeta(\redz) \eee \frac{18\pi^2+82[\Omega_\text{m}(\redz)-1]-39[\Omega_\text{m}(\redz)-1]^2}{\Omega_\text{m}(\redz)}$. We adopt the CDM subhalo mass-concentration relation \citep{Neto:2007vq}
\begin{eqnarray}\label{eqn:Subhalo_Mass_Concentration}
	\csub(\msub,\redz) = \frac{4.67}{1+\redz} \bigg(\frac{\msub}{10^{14} h_0^{-1}\text{M}_\odot}\bigg)^{-0.11},
\end{eqnarray}
as the FDM analogue is still largely unknown; here $h_0$ denotes the dimensionless Hubble constant. For instance, \citet{Du:2016zcv, Schutz:2020jox} instead adopt the warm DM mass-concentration relation. As noted later in Secs. \ref{sssec:SHMF_Tidal} and \ref{ssec:FDM_SHMF_Issues}, the mass accretion history of FDM subhaloes is physically distinct from those in CDM or warm DM cosmologies. The  $\redz$-dependence of $\csub$ suitable for FDM subhaloes may thus be fairly different from \eref{eqn:Subhalo_Mass_Concentration} as well as the warm DM counterpart. This particular modelling uncertainty and the adopted core-halo relation affect the detailed behaviour of tidal stripping. Nevertheless as discussed in Sec. \ref{ssec:FDM_Heating_Predictions}, the predicted heating is comparatively insensitive to these modelling choices.

The effect of tidal disruption becomes increasingly pronounced near the Galactic centre in suppressing the subhalo mass, quantified by the dimensionless truncation factor $\frac{T_R}{\msub}$. Following the analytical treatment in \citetalias{Church2019MNRAS}, we assume the leading-order subhalo mass loss occurs only outside the tidal radius
\begin{eqnarray}\label{eqn:Tidal_Radius}
	\rt = R\bigg[\frac{\msub}{2M_\text{enc}(R)}\bigg]^{1/3} \eee R_\text{vir}\bigg[\frac{\msub}{M_\text{enc}(R)}\bigg]^{1/3} \ft(R),
\end{eqnarray}
where $M_\text{enc}(R)$ denotes the total mass enclosed within $R$, and $\ft \eee \frac{R}{2^{1/3}R_\text{vir}}$. The adopted virial radius and concentration of the MW DM halo are $R_\text{vir} = 260$~kpc and $c= 15$ (see Sec. \ref{ssec:MW_Background_Disc_sigma_z}). 
The idealised simulations of \citet{Du:2018qor} find that naked subhalo soliton cores can experience complete tidal disruption within an orbital time if $\rt \lesssim \rc$. As the  FDM subhalo evolution in realistic cosmological simulations has not been studied, here we instead conservatively assume that the instantaneous tidal disruption of soliton cores occurs once $\rt\leq 3\rc$. Possible mismatches between this simplistic analytical treatment and realistic subhalo mass loss due to Galactic tides as well as relevant modelling uncertainties are discussed in Sec. \ref{sssec:SHMF_Tidal}.

Under a moderate assumption about the approximate self-similarity of subhaloes \citep{DeLucia:2003xe,Diemand:2007qr}, the subhalo scale radius can be expressed as
\begin{eqnarray}
	\rs(\msub,\redz,R) \simeq \frac{1}{\csub(\msub,\redz)}\frac{\rt(\msub,R)}{\ft(R)},
\end{eqnarray}
and the corresponding NFW profile extending from $3\rc$ to $\csub \rs$ has a mass scaled as $(\msub -M_\text{sol})\propto \log\big(\frac{1+\csub}{1+3\rc/\rs}\big) - \frac{\csub-3\rc/\rs}{(1+\csub)(1+3\rc/\rs)}$. 
The tidal truncation factor reads
\begin{eqnarray}\label{eqn:Truncation_Fun}
	\frac{T_R(\msub)}{\msub} =
	\begin{cases}
		0\;\;\;\;\;\;\;\;\;\;\;\;\;\;\;\;\;\;\;\;\;\;\;\;\;\;\;\;\:\:\:\text{(if $\rt\leq 3\rc$)},\\
		\frac{M_\text{sol}}{\msub}(1-\chi)+\chi \;\;\;\;\;\;\;\;\text{(if $\ft(R) <1$)},\\
		1\;\;\;\;\;\;\;\;\;\;\;\;\;\;\;\;\;\;\;\;\;\;\;\;\;\;\;\;\:\:\:\text{(otherwise)},
	\end{cases}
\end{eqnarray}
where the mass loss fraction in the truncated NFW profile is 
\begin{eqnarray}\label{eqn:Tidal_chi}
	\chi \eee \frac{\log\big(\frac{1+\rt/\rs}{1+3\rc/\rs}\big) - \frac{\rt/\rs-3\rc/\rs}{(1+\rt/\rs)(1+3\rc/\rs)}}{\log\big(\frac{1+\csub}{1+3\rc/\rs}\big) - \frac{\csub-3\rc/\rs}{(1+\csub)(1+3\rc/\rs)}}.
\end{eqnarray}
A vanishing truncation factor $\frac{T_R(\msub)}{\msub} = 0$ corresponds to the complete tidal disruption of a soliton-subhalo system.

The corresponding heating equation induced by subhalo transits can be computed by combining \pCAeref{eqn:Transient_Perturbation_Rate}\pCBeref{eqn:Tidal_Radius}\pCCeref{eqn:Truncation_Fun}
\begin{eqnarray}\label{eqn:Subhalo_Heating_Rate}
	\begin{cases}
		\bigH = \bigM\log\big(\frac{P}{\tau}\big),\\
		\bigM \eee \frac{8\pi G^2}{\sqrt{2}\sigmah}\int_{M_\text{sub,min}}^{M_\text{max}} d\msub \big[\msub\big(\frac{d \nsub}{d\log \msub}\big)\big(\frac{T_R}{\msub}\big)^2\big],\\
		\tau \eee \frac{2r_\text{min}}{\sigmah},
	\end{cases}
\end{eqnarray}
where $r_\text{min} \eee \min(\rt,\csub \rs)$ in the integrand denotes the subhalo termination radius after tidal stripping and carries an implicit $\msub$-dependence. \citetalias{Church2019MNRAS} set the integration upper bound to be $M_\text{max} = \Mh$ (i.e. the mass of host halo). However, $r_\text{min}$ increases with $\msub$ and could surpass the adiabatic cutoff $b_\text{max}$, resulting in a negative Coulomb logarithm. Hence the appropriate integration upper limit should be taken as either $\Mh$ or the subhalo mass that satisfies $r_\text{min}(\msub) = b_\text{max}$, whichever is smaller; namely, $M_\text{max}$ depends on both $\Mh$ and $b_\text{max}$. In evaluating \eref{eqn:Subhalo_Heating_Rate} we self-consistently update $r_\text{min}$ and $M_\text{max}$ at adequately small integration time step $\leq 0.05$~Gyr. Given the irreversibility of subhalo mass stripping, we adopt the truncation factor $T_R/\msub$ at redshift zero.

The tidal stripping treatment devised here is \textit{local}. Namely at each radius $R$, the tidal mass-loss model is applied to subhaloes of various masses. However the approximation does not account for subhalo populations with non-circular orbits that could experience different degrees of tidal stripping from subhaloes orbiting circularly at a fixed radius. Due to the dependence of subhalo destruction on the angular momentum (and not just the energy) of the subhaloes, there will be a Gaussian transition in effectiveness of subhalo-driven heating. We hence apply Gaussian kernel smoothing with a fixed width of $\pm1$~kpc to allow for the variation in the Coulomb logarithm cutoff due to angular momentum distribution.

\subsection{CDM Substructures: Differences and Similarities}\label{ssec:CDM_Counterparts}
In comparison to the FDM paradigm, the de Broglie-scale density granulation does not occur in CDM haloes. With the exact same background model, disc heating sourced by the baryon infall and accreted subhaloes is still captured by \pCAeref{eqn:Disc_Mass_Growth_Profile}\pCBeref{eqn:Heating_Eq_SGD_BGD_Limits}\pCCeref{eqn:Subhalo_Heating_Rate} with a modified subhalo distribution and tidal mass loss modelling. The CDM SHMF \eref{eqn:FDM_SHMF} is `unsuppressed' relative to the FDM counterpart and populates abundantly more low-mass subhaloes. Self-similar CDM subhaloes follow the NFW profile with the mass-concentration relation \eref{eqn:Subhalo_Mass_Concentration}, and the corresponding truncation factor reads
\begin{eqnarray}\label{eqn:CDM_Truncation_Fun}
	\frac{T_R(\msub)}{\msub} =
	\begin{cases}
		\frac{\log(1+\rt/\rs)-\frac{\rt/\rs}{1+\rt/\rs}}{\log(1+\csub)-\frac{\csub}{1+\csub}} \;\;\;\;\;\;\:\text{(if $\ft(R) <1$)},\\
		1\;\;\;\;\;\;\;\;\;\;\;\;\;\;\;\;\;\;\;\;\;\;\;\;\;\;\;\;\;\;\;\;\:\:\:\,\text{(otherwise)},
	\end{cases}
\end{eqnarray}
assuming that the fraction of CDM subhaloes undergoing complete tidal disruption is negligible \citep{vandenBosch2018MNRAS475}.

\section{Disc Heating Predictions and Constraints}\label{sec:Predictions_and_Results}

Section \ref{ssec:FDM_Heating_Predictions} compares recent measurements of the Galactic disc stellar kinematics with predicted disc heating caused respectively by FDM and CDM substructures and the resulting FDM particle mass constraint. We also point out that Gaia\textendash Enceladus and Sgr dwarf merger-driven perturbative heating alone unlikely accounts for the observed thick disc thickening in the solar vicinity. In Sec.~\ref{ssec:Sources_of_Error}, we carefully assess leading and subleading sources of error as well as modelling uncertainties in the empirical core-halo relation, FDM SHMF, and stellar heating estimate. The improvements and revised treatments compared to \citetalias{Church2019MNRAS} are discussed in Sec. \ref{sssec:Comparison_with_C2019}. We comment on the compatibility with and robustness of other relevant FDM constraints in Sec.~\ref{ssec:Literature_Comparison}. The possible physical origin of ultra-thin disc galaxies driven by anisotropic heating is explored in Sec. \ref{ssec:Anisotropic_Heating}.

\subsection{$\sigma_z$ Profiles and the Age\textendash Velocity Dispersion Relation}\label{ssec:FDM_Heating_Predictions}

The vertical velocity dispersion of the Galactic disc stars at birth is roughly $10$ km s$^{-1}$ \citep{Mackereth2019MNRAS, Miglio2021A&A645A}, consistent with the gaseous velocity dispersion \citep[e.g.][]{Malhotra1995ApJ448}. As the continuous mass infall \eref{eqn:Disc_Mass_Growth_Profile} and repeated transits of FDM substructures \pCAeref{eqn:Granulation_Heating_Rate}\pCCeref{eqn:Subhalo_Heating_Rate} drive the growth of $\sigma_z(R,t)$, we solve \eref{eqn:sigma_z_transition} and update the appropriate master equation \eref{eqn:Heating_Eq_SGD_BGD_Limits} in the SGD/BGD limit at sufficiently small time steps $\leq 0.05$ Gyr.

\subsubsection{Stellar Heating in the Solar Neighbourhood}\label{sssec:Solar_Neighbourhood}

The age\textendash velocity dispersion power law $\sigma_z \propto t^\beta$ in the solar vicinity has been commonly cited as a self-consistency test for various proposed disc heating models such as transient spiral modes \citep[e.g.][]{Lynden-Bell1972MNRAS157}, giant molecular cloud scattering \citep[e.g.][]{Lacey1984MNRAS208}, bar corotation \citep{Grand2016MNRAS459}, satellite interactions or accretion \citep[e.g.][]{Velazquez1999MNRAS304}, and interactions with MACHOs \citep{Lacey1985ApJ}. However, the power law index $\beta$ is still not tightly constrained, with observational values ranging from $0.25$ to $0.6$ (e.g. Table 1 of \citealp{Hanninen2002MNRAS337} and Table~8 of  \citealp{Sharma2014ApJ793}). Recent kinematic analyses based on large spectroscopic surveys like \textit{Gaia} and APOGEE also report statistically incompatible fits, with \citet{Mackereth2019MNRAS} finding $\beta_{\text{low-[$\alpha/$Fe]}}  = 0.5\pm 0.1$ and $\beta_{\text{high-[$\alpha/$Fe]}} = 0.02^{+0.12}_{-0.03}$, whereas \citet{Miglio2021A&A645A} show $\beta_{\text{low-[$\alpha/$Fe]}}  = 0.29\pm0.02$ and $\beta_{\text{high-[$\alpha/$Fe]}}  = 0.4\pm0.2$. \citet{Sharma2021MNRAS506} report $\beta_{\text{low-[$\alpha/$Fe]}} \simeq 0.26$ and that the age\textendash velocity dispersion relation for old populations deviates from a simple power law in the solar neighbourhood. The fitted power law index $\beta$ alone might not serve as a robust observational validation when assessing different disc heating mechanisms.

\begin{figure}
	\includegraphics[width=0.98\linewidth]{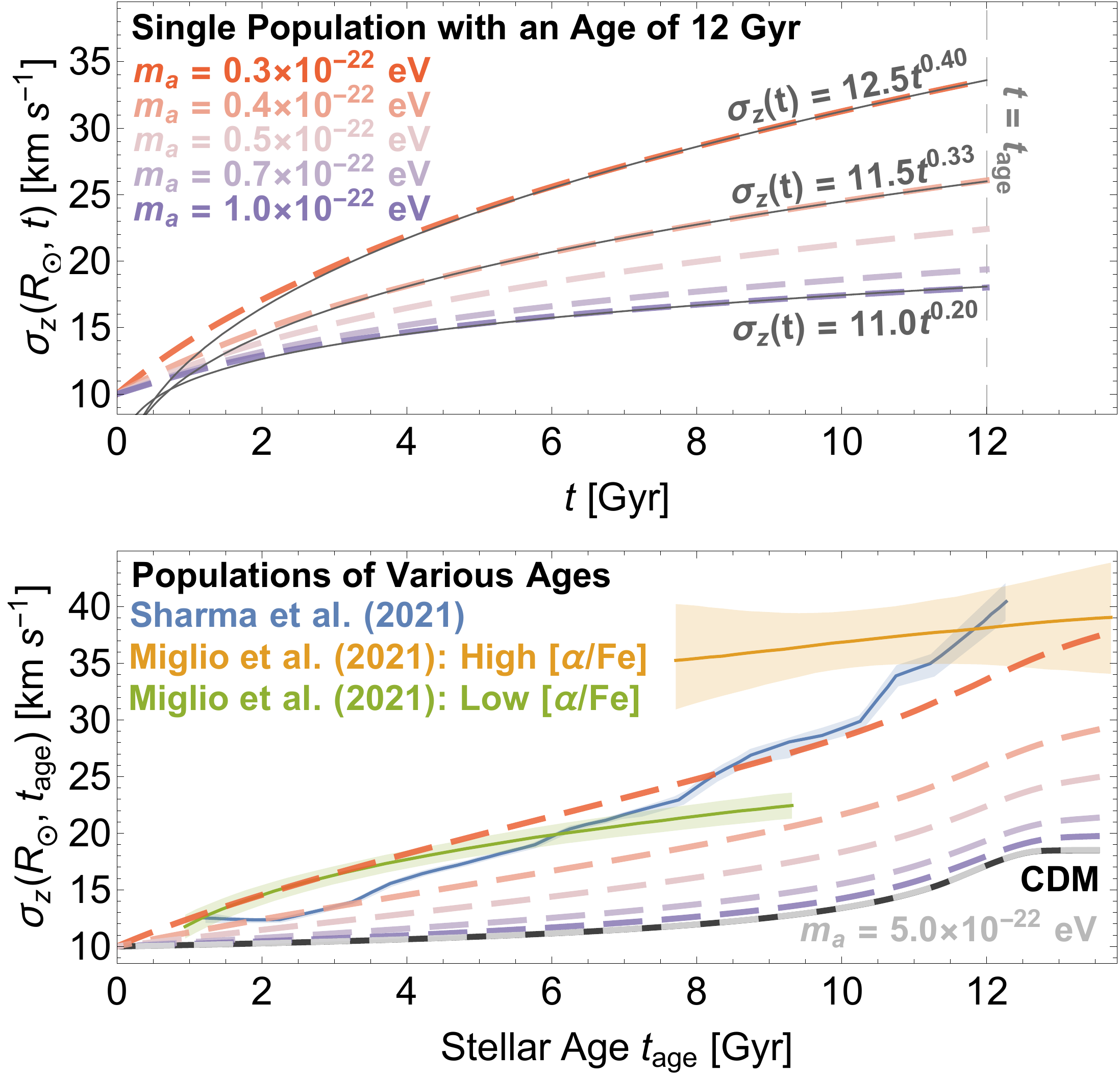}
	\caption{\textit{Top:} Time evolution of $\sigma_z$ for disc stars with $\tage=12$~Gyr in the solar neighbourhood appears well-fitted by the empirical scaling relation $\sigma_z \propto t^\beta$, fixing $m_a = 0.3$\textendash$1.0\times10^{-22}$~eV. \textit{Bottom:} Predicted vertical dispersion profiles (dashed) in the solar neighbourhood at the current epoch compared with the kinematic measurements by \citealp{Sharma2021MNRAS506} (blue) and \citealp{Miglio2021A&A645A} (yellow and green), parameterised by the stellar age $\tage$. The local heating signatures of FDM with $m_a = 5.0\times10^{-22}$~eV (grey) and CDM (black) are largely indistinguishable.}
	\label{fig:Age_Velocity_Dispersion}
\end{figure}

With this caveat in mind, we compare in the top panel of Fig.~\ref{fig:Age_Velocity_Dispersion} the age\textendash velocity dispersion curves at $R_\odot = 8.3$ kpc driven by FDM-substructure heating. A single population with an age of 12~Gyr is chosen to track $\sigma_z(R_\odot, t)$ \footnote{The time evolution of $\sigma_z(R_\odot, t)$ for a single stellar population with fixed $\tage$, strictly speaking, differs from the age\textendash velocity dispersion relation $\sigma_z(R_\odot, \tage)$ in observations that directly compares stellar populations with distinct $\tage$ at the \textit{current epoch}. Conventionally the former quantity is adopted in theoretical estimates (e.g. \citealp{Lacey1984MNRAS208, Lacey1985ApJ}; \citetalias{Church2019MNRAS}; \citealp{El-Zant:2019ios}). However the baryon mass infall onto the Galactic disc $d\Sigma/dt$ is strongly redshift-dependent (see Fig.~\ref{fig:MW-Disc_Mass_Infall}). The two definitions of $\sigma_z(R_\odot)$ can thus differ noticeably; cf. the two panels of Fig.~\ref{fig:Age_Velocity_Dispersion}.}; we have verified that the long-term heating behaviour does not vary strongly with the adopted stellar age. The best-fit power law index $\beta$ varies from 0.40\textendash0.20 for $m_a = 0.3$\textendash$1.0\times10^{-22}$~eV. Higher (lower) values of $\beta$ can be obtained for lighter (heavier) FDM particle masses. \citet{El-Zant:2019ios} found that disc heating caused by FDM halo granulation has a $m_a$-independent $\beta \simeq 0.5$. We note however that disc stars in the solar vicinity enter the SGD limit about 10 Gyr ago. The notable $m_a$-dependence in $\beta$ arises from the relative competition between the baryon infall-driven $d\Sigma/dt$ and FDM-induced stellar heating $\<\bigH\>$ as shown in \eref{eqn:Heating_Eq_SGD_BGD_Limits}. Another caveat about interpreting the power law index relates to the fact that old disc populations $\tage\gtrsim 8$ Gyr born in the inner disc might have experienced a rather distinct heating history before migrating to the solar neighbourhood $R_\odot$, a process not modelled in this work.

A physically more instructive assessment is to compare observational measurements directly with the predicted current-epoch $\sigma_z$ of disc populations born at different times. The bottom panel of Fig.~ \ref{fig:Age_Velocity_Dispersion} shows $\sigma_z(R_\odot,\tage)$ by \citet{Sharma2021MNRAS506} (blue; shading indicates the $68\%$ confidence interval) and \citet{Miglio2021A&A645A} (yellow for high-$[\alpha/\text{Fe}]$ populations and green for low-$[\alpha/\text{Fe}]$; shading indicates the respective 90$\%$ credible intervals). We additionally show the expected age\textendash velocity dispersion relations for FDM with $m_a = 5.0\times10^{-22}$~eV (grey) and CDM (black).

One of our main conclusions is that theoretical predictions with $m_a \simeq 0.3$\textendash$0.4\times 10^{-22}$~eV appear quantitatively consistent with the observed $\sigma_z(R_\odot,\tage)$ in the solar vicinity, given the adopted MW-sized DM halo profile with $\rhoh(R_\odot) = 0.0080$~M$_\odot$pc$^{-3}$. A factor-of-three uncertainty in the local DM density inference can nevertheless propagate to the preferred $m_a$, as discussed in Sec. \ref{sssec:Obs_Input_Stellar_Heating}. On the other hand, the observed age-chemo-kinematic properties indicate that the high-$\sigma_z$ tail of the age\textendash velocity dispersion comprises kinematically hot, high-[$\alpha$/Fe] stars with age $\gtrsim 8$~Gyr first formed in the inner disc region \citep[e.g.][]{Hayden2020MNRAS493, Sharma2021MNRAS506,Lu2022MNRAS512}. Hence focusing on only young disc populations that are less affected by radial migration, we require that the predicted heating for $\tage \lesssim 8$ Gyr does not exceed the observed $\sigma_z(R_\odot,\tage)$ to place a conservative FDM particle mass lower bound $m_a \gtrsim 0.4\times 10^{-22}$~eV. 

Lastly, it is worth highlighting that CDM and FDM with $m_a \gtrsim 5.0\times 10^{-22}$~eV give rise to indistinguishable stellar heating signatures in the solar neighbourhood. This is expected given the subhalo-driven disc thickening is subleading within $R\lesssim 10$~kpc, as later demonstrated in Fig.~\ref{fig:Heating_Decomp}. With the same background model, CDM and FDM behave comparably once the heating sourced by FDM granulation $\bigH \propto m_a^{-3}$ \eref{eqn:Heating_Time_Scale} subsides for sufficiently large $m_a$.

\subsubsection{Disc Vertical Velocity Dispersion Profile for $R \lesssim 20$ kpc}\label{sssec:Full_Disc_Profile}

\begin{figure}
	\includegraphics[width=0.98\linewidth]{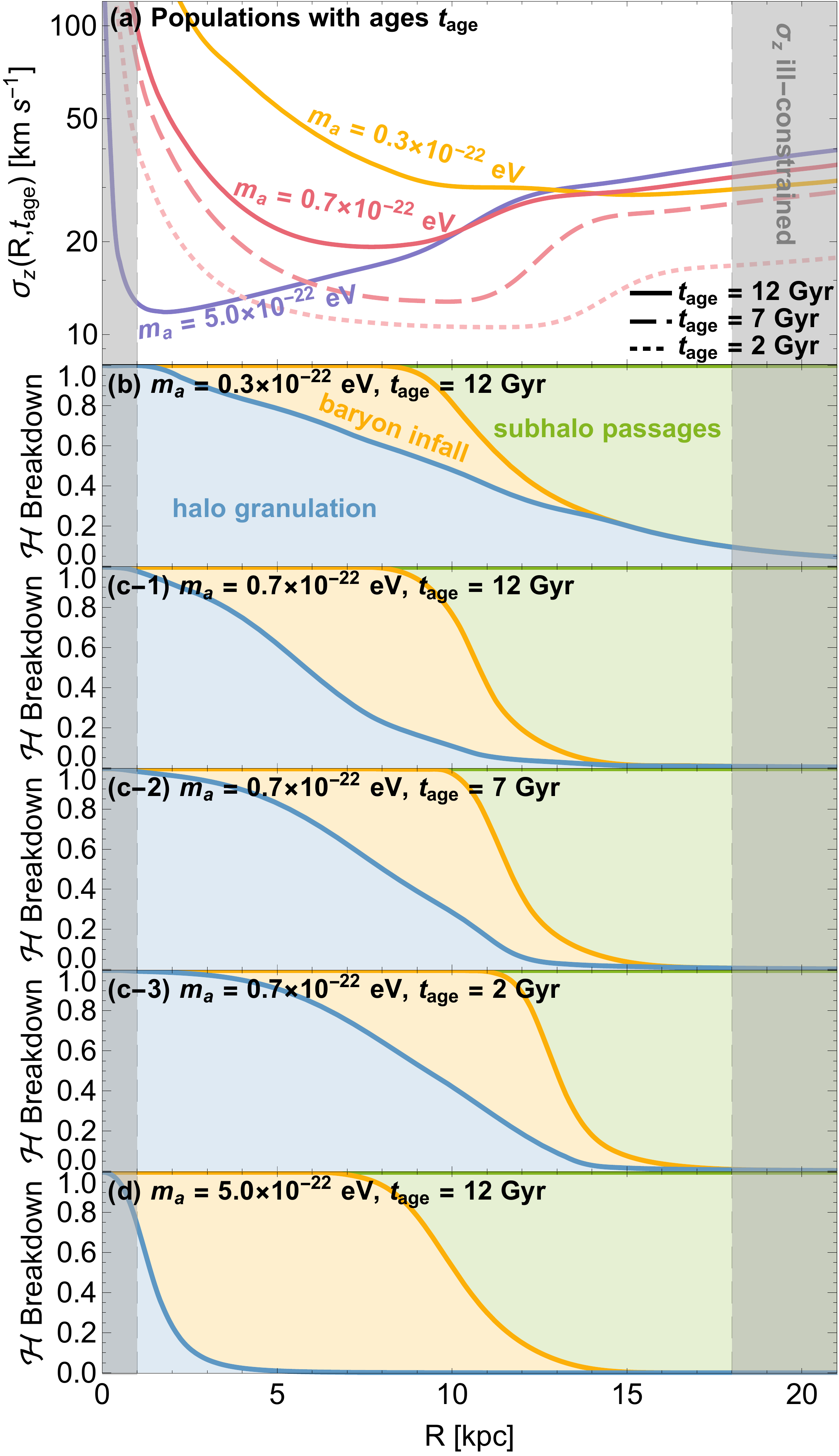}
	\caption{\textit{(a)} $\sigma_z(R,\tage)$ profiles for stellar populations of $\tage = 12, 7, 2$~Gyr (solid, dashed, dotted), fixing $m_a = 0.3, 0.7, 5.0\times10^{-22}$ eV (yellow, pink, and purple respectively). \textit{(b)-(d)} The relative contributions (normalised to unity) from halo granulation (blue shaded), baryon infall (yellow shaded), and subhalo passages (green shaded) to the total disc heating at each radius. The granule-driven heating is more prominent for smaller $m_a$ and dominates in the inner disc. In the outer disc region $R\gtrsim 10$~kpc, subhalo-induced potential perturbations become partially non-adiabatic $\ln\Lambda \geq 0$ and serve as the dominant \mbox{source of disc thickening, especially at later epochs and for larger $m_a$.}}
	\label{fig:Heating_Decomp}
\end{figure}

The $\sigma_z$ profile across the entire Galactic disc offers another observational test for the FDM heating mechanism. It is also instructive to assess the relative importance of each distinct mechanism discussed in Secs. \ref{ssec:Disc_Mass_Accretion} and \ref{sec:Dynamical_Evolution_Disc} at each radius $R$. Figure \hyperref[fig:Heating_Decomp]{6(a)} shows the predicted $\sigma_z(R)$ profiles for stellar populations with ages of $\tage = 12$~Gyr (solid), $7$~Gyr (dashed), and $2$~Gyr (dotted), fixing $m_a = 0.3, 0.7, 5.0\times10^{-22}$~eV (yellow, pink and, purple respectively). Panels \hyperref[fig:Heating_Decomp]{6(b)}, \hyperref[fig:Heating_Decomp]{6(c-1)}, and \hyperref[fig:Heating_Decomp]{6(d)} compare the breakdown $\Delta \sigma_{z,i}^2/\Delta \sigma_{z,\text{tot}}^2$ of total disc heating $\bigH$ sourced by halo granulation (blue shaded), baryon mass infall (yellow shaded), and orbiting subhaloes (green shaded), fixing $\tage = 12$ Gyr. We  fix $m_a = 0.7\times 10^{-22}$~eV and adopt $\tage = 12, 7, 2$~Gyr in panels \hyperref[fig:Heating_Decomp]{6(c-1)}, \hyperref[fig:Heating_Decomp]{6(c-2)}, and \hyperref[fig:Heating_Decomp]{6(c-3)} respectively.

The effect of time-varying density granulation \eref{eqn:Granulation_Heating_Rate} $\propto m_a^{-3}\rhoh^2$ is most pronounced in the inner disc and for lighter $m_a$, as evident from panels \hyperref[fig:Heating_Decomp]{6(b)}, \hyperref[fig:Heating_Decomp]{6(c-1)}, and \hyperref[fig:Heating_Decomp]{6(d)}. The Galactic disc at the current epoch appears self-gravitating for $1$~kpc~$\lesssim R\lesssim 12$~kpc (Fig. \ref{fig:MW-Disc_sigma_z-h(R)}). For disc stars that have undergone (nearly) self-gravitating phases, continuous baryon infall can contribute non-trivially to the total disc thickening and is more effective with increasing redshift (Fig. \ref{fig:MW-Disc_Mass_Infall}). 

Subhalo-driven heating expectedly dominates in the outer disc region as the extent of Galactic tidal stripping wanes, irrespective of the adopted $m_a$ and stellar age $\tage$. This heating source ceases entirely within $R \lesssim 7$ kpc. At a fixed radius $R$, the disc self-gravity increases with time (lower redshift) due to mass infall, simultaneously lowering the disc star vertical oscillation periods in the SGD limit \eref{eqn:Disc_Osc_Period} and the value of adiabatic cutoff $b_\text{max}$ (see the discussion around \eref{eqn:Subhalo_Heating_Rate}). Gravitational perturbations caused by subhalo passages become entirely adiabatic to a stellar population if $\ln\Lambda \leq 0$. Since weaker total vertical gravity translates to longer vertical oscillation timescale and hence larger $b_\text{max}$, the Coulomb factor $\Lambda = b_\text{max}/b_\text{min}$ increases with larger $R$. Note that young, thin disc stars born kinematically cold could still be self-gravitating even beyond $R \gtrsim 12$~kpc. The transition from $\ln\Lambda \leq 0$ to $\ln\Lambda \geq 0$ takes place at progressively larger radii for younger populations. This short period of rapid increase in $\ln\Lambda$ after zero-crossing, followed by a gentler logarithmic growth, is imprinted in the predicted $\sigma_z(R)$ profiles. Furthermore, this heating channel is mainly sourced by the massive tail of subhalo populations. The resultant dynamical effects are thus largely unaffected by the $m_a$-dependent low-mass cutoff in FDM SHMF and detailed tidal stripping prescription (see also Sec. \ref{sssec:SHMF_Tidal}). We expectedly observe that the subhalo-induced heating is comparatively insensitive to the adopted $m_a$ across $0.3$\textendash$5\times 10^{-22}$~eV.

\begin{figure}
	\includegraphics[width=0.98\linewidth]{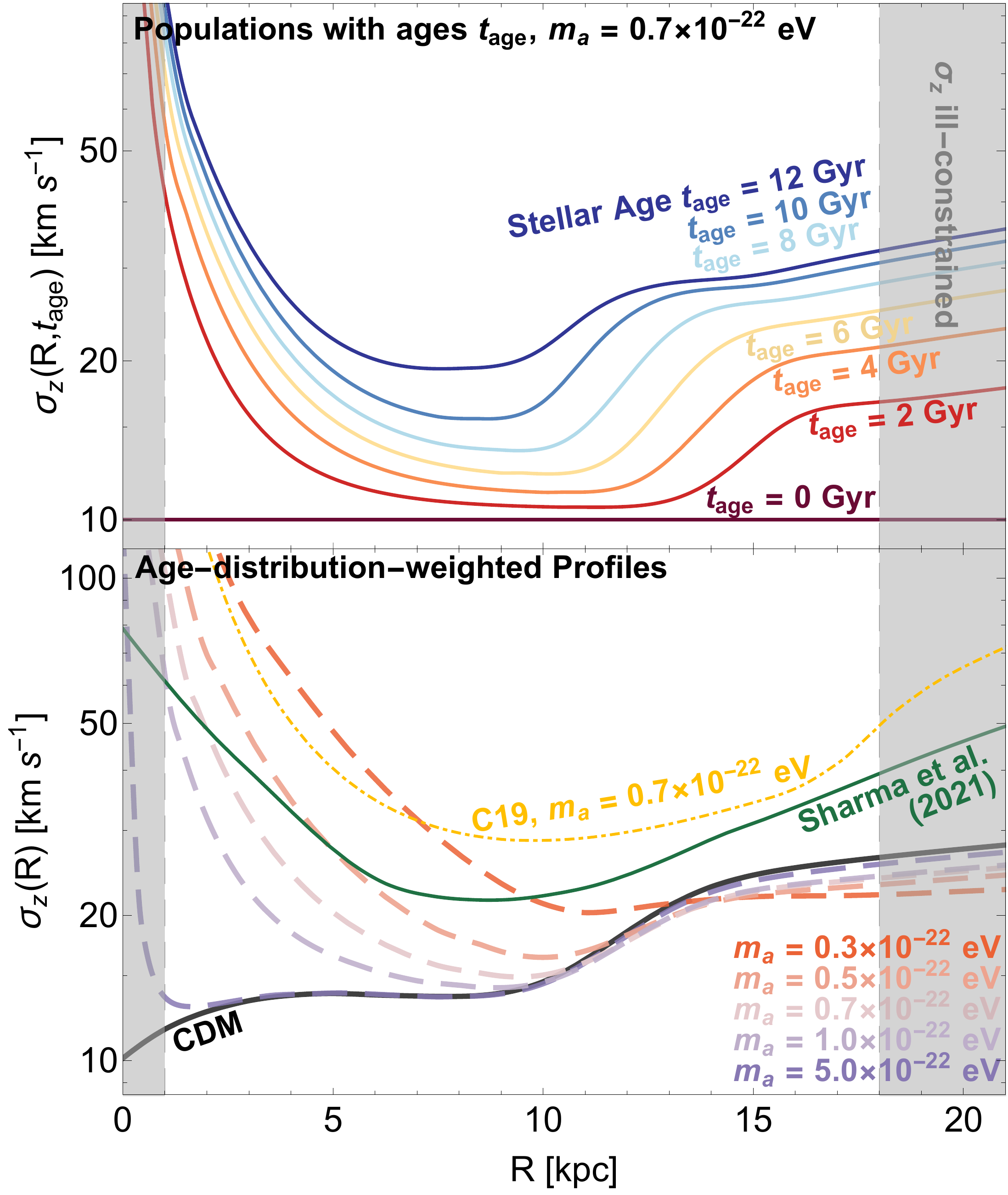}
	\caption{\textit{Top:} $\sigma_z(R)$ profiles for stellar populations of $t = 0$\textendash$12$~Gyr, fixing $m_a = 0.7\times 10^{-22}$ eV; the same as the top panel of Fig. \ref{fig:Heating_Decomp} with more stellar populations. The heating in inner and outer disc regions are primarily sourced by halo granulation and orbiting subhaloes respectively, bridged by the baryon infall-driven heating for $7$~kpc~$\lesssim R\lesssim12$~kpc. This non-uniform, $R$-dependent heating rate results in an overall U-shaped pattern. \textit{Bottom:} Predicted age-distribution-weighted dispersion profiles for FDM with $m_a = 0.3$\textendash$5.0\times 10^{-22}$~eV (dashed) and CDM (black solid) compared with that of \citetalias{Church2019MNRAS} fixing $0.7\times 10^{-22}$~eV (dot-dashed yellow) and the best-fit model by \citet{Sharma2021MNRAS506} (solid green) from analyzing $\simeq 840,000$ stellar samples.}
	\label{fig:Age_Weighted_Profiles}
\end{figure}

The top panel of Fig.~\ref{fig:Age_Weighted_Profiles} shows the predicted vertical velocity dispersion profiles for stellar populations with ages of $\tage = 0$\textendash12~Gyr, fixing $m_a = 0.7\times 10^{-22}$~eV. Potential fluctuations caused by FDM wave interference and continuous baryon infall (in the SGD limit) are responsible for the salient disc thickening observed in the inner disc region. Stellar heating driven by repeated subhalo passages rises gently in effectiveness at larger radii, as tidal stripping of $\msub$ subsides. This non-uniform heating signature reaches the global minimum at around $R \simeq 8$\textendash12~Gyr, \mbox{depending on the stellar age $\tage$.}

The inside-out mass growth pattern introduces a strong radial dependence in the ensemble-averaged disc stellar age, as discussed in Sec. \ref{ssec:Disc_Mass_Accretion} (see also Fig. \ref{fig:Disc_Age_Distribution}). We therefore assign an appropriate weighting to $\sigma_z(R,\tage)$ based on the reconstructed stellar age distribution. The bottom panel of Fig. \ref{fig:Age_Weighted_Profiles} compares the predicted age-distribution-weighted $\sigma_z$ profiles (dashed) with the weighted best-fit model by \citet{Sharma2021MNRAS506} (solid green; see Sec. \ref{ssec:MW_Background_Disc_sigma_z} and Fig. \ref{fig:MW-Disc_sigma_z-h(R)}). In comparison, the heating signature for $m_a = 0.7\times 10^{-22}$ eV predicted by \citetalias{Church2019MNRAS} (dot-dashed yellow) assumes a uniform heating time (stellar age) of $12$ Gyr and being self-gravitating across the entire Galactic disc (see Sec. \ref{sssec:Comparison_with_C2019} for further details).

The predicted U-shaped disc heating signature is qualitatively consistent with the disc kinematic measurements, favouring $m_a = 0.5$\textendash$0.7\times 10^{-22}$~eV. The excess of stellar heating predicted for $R\lesssim 3$ kpc partly results from the inner DM density overestimated by inward extrapolating a cuspy NFW profile. In particular for $m_a = 5.0\times 10^{-22}$~eV, the granulation-induced stellar heating $\propto m_a^{-3} \rhoh^2$ is expected to be largely negligible. Yet overheating is still present near the Galactic centre $R \ll 1$~kpc, aside from the minimal stellar heating $\Delta \sigma_z \simeq 3$~km~s$^{-1}$ due to baryon infall. We note also that the region within $R \lesssim 1$~kpc is dominated not by the disc but by the bulge, and the large velocity dispersion $\gtrsim 100$~km~s$^{-1}$ is not inappropriate for that component \citep[e.g.][]{Valenti2016AA587L}. Repeated gravitational encounters with orbiting subhaloes predicts a nearly constant $\sigma_z$ profile in the outer disc region $R \gtrsim 15$~kpc. The discrepancy with the best-fit model by \citet{Sharma2021MNRAS506} may be reduced by also considering heating contributions from minor mergers effective at large $R$ \citep{Quinn1993ApJ403}.

The resolved age-chemical abundance structures of stellar disc populations provide strong observational evidence of efficient radial migration $\Delta R\simeq 4$\textendash$6$ kpc over the formation timescale \citep[e.g.][]{Frankel2019ApJ884F,Miglio2021A&A645A,Lu2022MNRAS512}. Significant orbit mixing (i.e. both inwards and outwards migration) acts as a global diffusion process that smooths out the uneven heating signature \citep{Frankel2018ApJ865,Okalidis2022MNRAStmp}. In particular, the inner `spike' of the U-shaped heating signature should be less pronounced, exchanging some high-$\sigma_z$ stars with kinematically cold populations born in the intermediate region $6\text{ kpc}\lesssim R \lesssim 10$ kpc over $\gtrsim 8$ Gyr. Incorporating the effects of a typical radial diffusion and migration process as well as other conventional stellar heating mechanisms would increase the total stellar heating rate in the solar neighbourhood, hence pushing the favoured FDM particle to $m_a \gtrsim 0.7\times 10^{-22}$~eV. Self-consistently coupling the present heating mechanism to an analytical radial diffusion model is beyond the scope of this work.

\subsubsection{Gaia\textendash Enceladus and Sgr Dwarf Merger-driven Disc Heating}

The predicted disc thickening for CDM and FDM (Fig. \ref{fig:Age_Weighted_Profiles}) differs most noticeably within the solar radius $\lesssim 8$~kpc and becomes compatible for $m_a \gtrsim 5\times10^{-22}$~eV, given the strong scaling $T_\text{heat}^{-1} \propto m_a^{-3}$ in the granulation-driven stellar heating \eref{eqn:Heating_Time_Scale}. For FDM with $m_a \geq 1.0\times10^{-21}$~eV and CDM, the entire age-distribution-weighted dispersion profiles between $1$~kpc~$\leq R\leq$~20~kpc agree within $\leq1\%$ and are indistinguishable from current observations. One might naturally ask in the CDM paradigm whether there could already be sufficient inner disc heating caused by the now well-documented Gaia\textendash Enceladus accretion event taking place $\simeq 10$~Gyr ago and the ongoing merger of the Sgr dwarf. If so, we first note that the viable FDM particle bound inferred from the local disc star kinematics (bottom panel of Fig.~\ref{fig:Age_Velocity_Dispersion}) would be pushed to $m_a \gtrsim 5\times10^{-22}$~eV, instead of ruling out the entirety of possible FDM particle masses. Also, the disc thickening due to continuous mass accretion per se has been accounted for by the $d\log\Sigma/dt$ term in \eref{eqn:Heating_Eq_SGD_BGD_Limits}.

Here we point out that these two mergers unlikely serve as the sole dominant perturbative heating mechanism to produce the observed thick disc kinematics in the solar vicinity. First, the current Sgr dwarf merger is believed to provide a possibly important source of disc thickening \citep[e.g.][]{Ruiz-Lara2020NatAs4965R}. However, recent simulations carried out by \citet{Bennett2022ApJ927131B} demonstrate that for Sgr dwarf sufficiently massive to cause the observed mean vertical velocity perturbations in the solar neighbourhood, the resulting vertical number count asymmetry is grossly incompatible in both amplitude and wavelength with the \textit{Gaia} DR2 data. They confidently rule out interactions with Sgr as the dominant cause of density waves in the Galactic disc over a realistic range of Sgr progenitor masses, MW models, and merger initial conditions. The Sgr-driven perturbative heating in the solar neighbourhood is thus expected to be sub-leading.

Simulations of Gaia\textendash Enceladus and MW analogues show that (1)~the~merger of Gaia\textendash Enceladus progenitor, if massive enough, heats most already-formed disc stars by up to a factor of two in $\sigma_z$, (2) the accretion-driven dynamical impact on disc kinematics is delivered within $\lesssim 1$~Gyr, and (3) the level of kinematic heating is sensitive to the Gaia\textendash Enceladus progenitor mass \citep{Bignone2019ApJ883L5B, Grand2020MNRAS4971603G}. The progenitor mass is currently not well-constrained, with values ranging from $< 10^9$~M$_\odot$ \citep{Mackereth2019MNRAS4823426M, Grand2020MNRAS4971603G} to $\sim 5\times 10^9$~M$_\odot$ \citep{Helmi2018Natur56385H, Vincenzo2019MNRAS487L47V}. With a median stellar age $12.33_{-1.36}^{+0.92}$~Gyr, the inferred chemical evolution of Gaia\textendash Enceladus indicates that the merger took place $\simeq 10$~Gyr ago \citep{Helmi2018Natur56385H, Vincenzo2019MNRAS487L47V, Montalban2021NatAs5640M}. If the major merger is solely responsible for establishing the present-day thick disc kinematics, populations formed prior to and post this violent episode of perturbative heating should exhibit a clear jump in $\sigma_z(\tage)$. Namely the thick disc age\textendash velocity dispersion relation should appear comparatively flat on both ends and connected by a short period of merger-driven rapid increase around $\tage\simeq 10$~Gyr, which is not obvious in the data of \citet{Sharma2021MNRAS506} and absent in that of \citet{Miglio2021A&A645A} for the local high-[$\alpha/$Fe] populations (Fig.~\ref{fig:Age_Velocity_Dispersion}). Identifying such an unmistakable kinematic signature in the inner thick disc age\textendash velocity dispersion relation could better assess the dynamical impact of Gaia\textendash Enceladus merger across the entire Galactic disc.

\subsection{Uncertainties in FDM Predictions and Sources of Error}\label{ssec:Sources_of_Error}

\subsubsection{Core-halo Relation in the MW}\label{sssec:Core-halo_Soliton_Size}

In a sequence of idealised mergers of soliton-halo systems, linear growth $M_\text{sol} \propto \Mh^\alpha$ with $\alpha = 1$ is expected if the sum of soliton masses is conserved. However the soliton effective temperature (specific energy) would then have increased more rapidly than that of the host halo; FDM haloes with sufficiently high thermal conductivity should instead give rise to a gentler slope  \mbox{$0<\alpha<1$} \citep{Schive:2014hza,Hui:2016ltb,Bar:2018acw, Bar:2019bqz}. \citet{Schive:2014hza} observe $\alpha = 1/3$, \eref{eqn:Core_Halo_Relation}, in relaxed FDM haloes $\Mh \lesssim 4\times10^{11}$~M$_\odot$ from both cosmological simulations, with $m_a = 0.8\times 10^{-22}$~eV and box sizes $2, 20, 40$~Mpc, and dimensionless soliton mergers simulations. The result is independently confirmed by \citet{Veltmaat:2018dfz} with cosmological simulations reaching $\Mh\leq 7\times 10^{10}$~M$_\odot$, adopting $m_a = 2.5\times 10^{-22}$~eV and a box size $\simeq 3.6$~Mpc, and \citet{May:2021wwp} with large-box (14 Mpc) cosmological simulations reaching $\Mh \lesssim 10^{10}$~M$_\odot$, fixing $m_a = 0.7\times 10^{-22}$~eV. The statistical spread in $M_\text{sol} \propto \Mh^{1/3}$ reported by \citet{Schive:2014hza} and \citet{May:2021wwp} is about a factor of~$2$ \mbox{(e.g. see Fig. 5 of \citet{Chan:2021bja}).}

However the uniqueness of this empirical correlation has been disputed in some other numerical studies. \citet{Mocz:2017wlg} find $\alpha = 5/9$ from soliton-merger simulations, having $m_a = 2.5\times 10^{-22}$~eV and a box size $\simeq 1.8$~Mpc. Simulated FDM haloes, albeit with a limited sample size, in cosmological simulations of \citet{Mina:2020eik} also appear consistent with $\alpha = 5/9$, fixing $m_a = 2.5\times 10^{-22}$~eV and a box size $\simeq 3.7$~Mpc. \citet{Chan:2021bja} perform both soliton-merger and small-box ($0.34$~Mpc) cosmological simulations with $m_a = 1.0\times 10^{-22}$~eV and find $\alpha \simeq 1/3$\textendash$0.9$. \citet{Schwabe:2016rze} observe a significant spread in the $M_\text{sol}$\textendash$\Mh$ correlation \footnote{\label{fn:FDM_N-body} This large spread is similarly reported in \citet{Nori:2020jzx}, where eight FDM haloes are simulated with the $N$-body {\small AX-GADGET} code \citep{Nori:2018hud}. We caution that the smoothed particle hydrodynamics numerical methods employed therein has thus far not been verified for the accuracy in nonlinear regimes where vortices can develop along the zero-density voids. Relatedly, the analytical estimate by \citet{Taruya:2022zmt} suggests that the observed scatter in the $M_\text{sol}$\textendash$\Mh$ correlation might be attributed to the intrinsic dispersion in the halo mass-concentration relation.} in soliton merger simulations, adopting $m_a = 2.5\times 10^{-22}$~eV and box sizes $\simeq 1$~Mpc. Possible culprits include different simulation setups, spatial resolution limit, small box sizes, merger history, and tidal mass loss of $\msub$ \citep{Chan:2021bja}. Future numerical studies that systematically address these physical or artefactual factors could help clarify the $M_\text{sol}$\textendash$\msub$ correlation with greater certainty. Massive central baryon distribution could also impact the core-halo relation found in pure-FDM simulations; we leave the investigation to future work.

Regardless whether $\alpha = 1/3$ is assumed, the soliton core density $\rhoc$ and $\rc$ are still related via the scaling symmetry of the Schr\"{o}dinger\textendash Poisson equations \citep{Schive:2014dra}
\begin{eqnarray}\label{eqn:rc_Free}
	\rc(m_a,\rhoc) = 0.37\bigg(\frac{\rhoc}{\text{ M}_\odot\text{pc}^{-3}}\bigg)^{-1/4}\bigg(\frac{m_a}{10^{-22}\text{ eV}}\bigg)^{-1/2}\text{ kpc}.
\end{eqnarray}
As a conservative estimate, we assume a factor-two statistical spread both above and below the empirical relation  $M_\text{sol} \propto \Mh^{1/3}$. In an MW-sized host halo $\Mh = 10^{12}$~M$_\odot$, adopting $m_a = 0.3(1.0)\times10^{-22}$~eV translates to bounds on the core radius $0.27\text{ kpc}\leq \rc \leq 1.1$~kpc ($0.079\text{ kpc}\leq \rc \leq 0.32$~kpc) for $\alpha = 1/3$; higher power law index $\alpha \geq 1/3$ predicts cores with increasingly smaller $\rc$. The MW total baryon mass $\simeq 8\times 10^9$~M$_\odot$ enclosed within $r = 1$~kpc (Fig.~\ref{fig:MW-Total}) is comparable or more massive than $M_\text{sol}$ considered here. For a fixed $M_\text{sol}$, the gravitational potential well generated by such a non-trivial central baryon distribution can distort the soliton density profile, reducing $\rc$ by a factor of $\simeq 2$ when $M_\text{sol} \simeq M_\text{baryon}$ within 3$\rc$ \citep{Bar-Or2019ApJ}. The confined soliton random-walk excursions observed in $\redz = 0$ simulations by \citet{Schive:2019rrw} and \citet{Chowdhury2021ApJ} can efficiently heat central baryon structures within a few Gyr. We conclude that for the $m_a$ range of interest, dynamical effect of the Galactic soliton is largely uncertain for $R \gtrsim 1$~kpc and hence not modelled in this work.

\subsubsection{FDM Subhalo Mass Function and Tidal Stripping Modelling}\label{sssec:SHMF_Tidal}

Here we examine the sizeable discrepancy amongst simulation-inferred and analytical FDM SHMFs. \citet{Schive:2015kza} provide the first numerical estimate of FDM SHMF \pCAeref{eqn:FDM_SHMF}\pCCeref{eqn:M0_FDM_SHMF_Schive} for $\redz = 4$\textendash$10$, fixing $m_a = 0.8$\textendash$3.2\times 10^{-22}$~eV. With initial conditions constructed from an analytical FDM transfer function that could mildly underestimate low-mass subhaloes, the authors conduct collisionless $N$-body simulations and later remove spurious low-mass subhaloes. 

\citet{May:2021wwp} report the first FDM SHMF extracted from self-consistent FDM cosmological simulations fixing $m_a = 0.35\times 10^{-22}$ and $0.7\times 10^{-22}$ eV, although the adopted CDM initial conditions overpopulate low-mass subhaloes. Their resulting FDM SHMFs at $\redz = 3$ are well parameterised by \eref{eqn:FDM_SHMF} with $\alphasub = -0.64, \betasub = -2.2$, and a suppression mass scale \footnote{We thank Simon May for clarifying the typo in the reported $M_0(m_a)$; see the legend of Fig.~12 in \citet{May:2021wwp}.}
	\begin{eqnarray}\label{eqn:M0_FDM_SHMF_May}
		M_0(m_a) = 9.1\times 10^8 \bigg(\frac{m_a}{10^{-22}\text{ eV}}\bigg)^{-4/3}\text{ M}_\odot,
	\end{eqnarray}
lower by a factor of $\simeq10^{1.3}$ compared with \eref{eqn:M0_FDM_SHMF_Schive}.

However in their follow-up FDM cosmological simulations with FDM initial conditions, \citet{May2022arXiv220914886M} find that the FDM SHFM at $\redz = 3$ for $m_a = 0.7\times 10^{-22}$ eV is largely consistent with \pCAeref{eqn:FDM_SHMF}\pCCeref{eqn:M0_FDM_SHMF_Schive} estimated by \citet{Schive:2015kza}. The non-trivial discrepancy between  \pCAeref{eqn:M0_FDM_SHMF_Schive}\pCCeref{eqn:M0_FDM_SHMF_May} is attributed to the adopted FDM versus CDM initial power spectra. Limited by the small number of simulated low-mass haloes even down to $\redz = 3$, they caution that the conclusive corroboration of \pCAeref{eqn:FDM_SHMF}\pCCeref{eqn:M0_FDM_SHMF_Schive} at other redshifts requires future FDM cosmological simulations with even larger box sizes.

\citet{Du:2016zcv} provide a semi-analytical estimate of FDM SHMF that has enjoyed widespread popularity with works aiming to constrain $m_a$ (e.g. \citealp{Marsh:2018zyw,Schutz:2020jox,Banik:2019smi,DES:2020fxi}; see Sec. \ref{ssec:FDM_SHMF_Issues}). Based on the extended Press\textendash Schechter formalism, the approach adopts the empirical core-halo relation \eref{eqn:Core_Halo_Relation} (see Sec. \ref{sssec:Core-halo_Soliton_Size} for related uncertainties), an explicit warm DM tidal stripping prescription, and warm DM mass-concentration relation, as the FDM analogues are still largely unknown. However, the nonlinear dynamics and evolution of warm DM and FDM are quantitatively different, due to the distinct physical origins \citep{Mocz:2019pyf,DES:2020fxi}. Furthermore, their use of a spherical top hat, as opposed to the sharp-$k$ filter, window function and the mass-dependent critical density barrier function can render the calculation self-inconsistent \citep{Kulkarni:2020pnb}. 

\begin{figure}
	\centering	
	\includegraphics[width=0.98\linewidth]{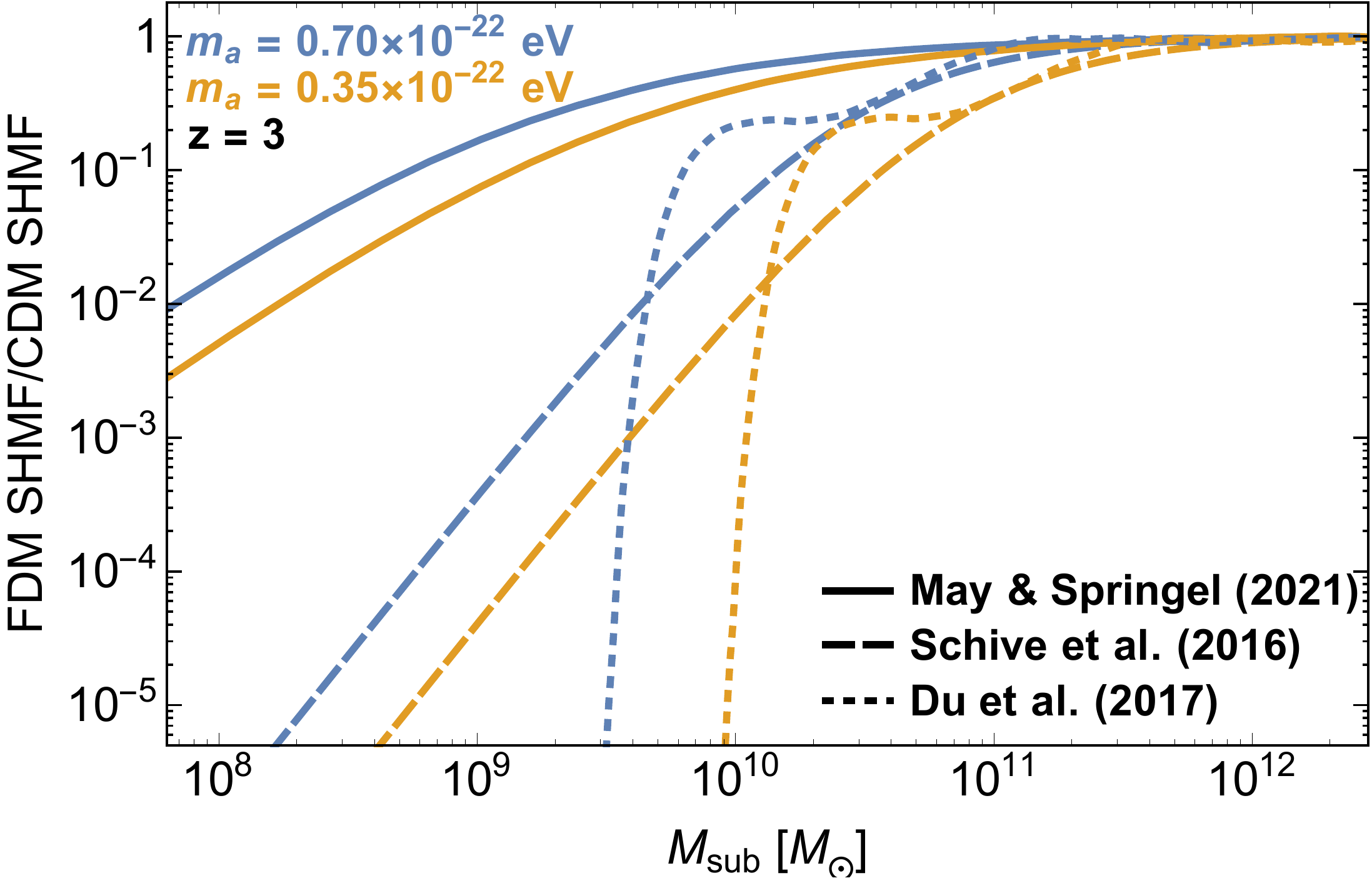}
	\caption{Suppression of unevolved (i.e. pre-infall) FDM SHMF relative to CDM SHMF $\frac{d \nsub}{d \msub}\big|_\text{FDM}/\frac{d \nsub}{d \msub}\big|_\text{CDM}$ defined in \eref{eqn:FDM_SHMF}, observed in \citet{May:2021wwp} (solid), \citet{Schive:2015kza} (dashed), and predicted by the semi-analytical treatment of \citet{Du:2016zcv} (dotted) at redshift $\redz = 3$, fixing $m_a = 0.7\times 10^{-22}$ eV (blue) and $0.35\times 10^{-22}$ eV (yellow).}
	\label{fig:FDM_SHMF}
\end{figure}

Figure \ref{fig:FDM_SHMF} compares the relative suppression between the unevolved (prior to accretion) FDM and CDM SHMFs observed in \citet{May:2021wwp} (solid; \eref{eqn:M0_FDM_SHMF_May}), \citet{Schive:2015kza} (dashed; \eref{eqn:M0_FDM_SHMF_Schive}), and predicted by the semi-analytical prescription of \citet{Du:2016zcv} (dotted), fixing the redshift $\redz = 3$ and $m_a = 0.7\times 10^{-22}$~eV (blue), $0.35\times 10^{-22}$~eV (yellow). There appears to be \textit{no overall quantitative agreement.} Discrepancy between the former two empirical relations again stems from the adopted CDM or FDM initial power spectra. \citet{Du:2016zcv} predict sharp low-mass cutoffs, qualitatively inconsistent with simulations. For $m_a = 0.7\times10^{-22}$~eV, their estimated minimal halo mass $3\times 10^9$~M$_\odot$ is also incompatible with $M_\text{sub,min} \simeq 2\times 10^8$~M$_\odot$ found in FDM simulations \eref{eqn:Core_Halo_Relation}. The analytical estimate by \citet{Kulkarni:2020pnb} similarly gives a comparative dearth of low-mass subhaloes. Provided that analytical SHMFs all predict suppression behaviours dissimilar to numerical counterparts, \citet{May:2021wwp} argue analytical simplifications that rely on Jeans filtering to suppress small-scale structures might not accurately capture low-redshift nonlinear dynamics in an FDM cosmology.

This modelling uncertainty becomes progressively acute for lighter $\Mh$ and $m_a$.  The FDM SHFM for $m_a \simeq 10^{-22}$~eV is highly uncertain below $\msub \leq \bigO(10^{9})$ M$_\odot$ (Fig. \ref{fig:FDM_SHMF}). The subhalo-induced heating \eref{eqn:Subhalo_Heating_Rate} $\propto\msub(d\nsub/d\log\msub)$ \footnote{\label{fn:CDM_SHMF} The unevolved CDM SHMF (and hence the high-mass tail of FDM SHMF) roughly follows $d\nsub/d\log\msub\propto \msub^{-0.9}$ at $\redz = 0$ \citep{Han:2015pua}.} receives greater contributions from massive subhaloes and therefore is comparatively insensitive to the precise distribution of low-mass subhalo populations. In our tidal stripping prescription (Sec. \ref{ssec:FDM_Subhalo_Passages}), subhaloes with \textit{initial} masses $M_\text{sub,i} \lesssim 8(3)\times10^9$ M$_\odot$ are completely disrupted by Galactic tides within $R\simeq 8(18)$ kpc for $m_a \simeq 10^{-22}$~eV. The Galactic disc thickening driven by subhalo passages is negligible within the solar radius and dominates only at large radii (Fig. \ref{fig:Heating_Decomp}) where tidal stripping is less severe. We have numerically verified that adopting different FDM SHFM parameterisations of \citet{Schive:2015kza,  Du:2016zcv, May:2021wwp} incurs only subleading corrections $|\Delta\sigma_z(R)|/\sigma_z(R) \simeq 0.1 (\simeq 0.01)$ to the predicted subhalo-induced heating for $14$~kpc~$\lesssim R\lesssim$~20~kpc, fixing $m_a = 0.5(5.0)\times 10^{-22}$ eV. This heating channel is hence robust and comparatively unaffected by different SHFM modelling choices.

Next we assess the tidal stripping model introduced in Sec.~\ref{ssec:FDM_Subhalo_Passages}, which accounts for only subhalo mass loss lying outside the tidal radius $\rt$ and can underestimate the tidal effects. \citet{Schive:2019rrw} find that an FDM subhalo $M_\text{sub,i} \simeq 6\times 10^9$~M$_\odot$ with $r_\text{vir} \simeq 50$~kpc, $M_\text{sol} \simeq 10^8$~M$_\odot$, and $\rc \simeq 0.7$~kpc on a circular orbit of $R = 100$ kpc around an MW-mass point mass experiences significant mass loss $M_\text{sub,f}/M_\text{sub,i}\simeq 0.1$ within $\sim$ 1~Gyr. In contrast, a similar level of subhalo mass loss $\frac{T_R(\msub)}{\msub} = 0.1$ quantified by \eref{eqn:Truncation_Fun} takes place on a much closer orbit $R \simeq 25$~kpc. This non-trivial discrepancy could stem from the inapplicability of CDM mass-concentration relation \eref{eqn:Subhalo_Mass_Concentration} to low-mass FDM subhaloes that are comparatively more vulnerable to tidal stripping, on top of the equivocal nature of suitable tidal radius definitions \citep{vandenBosch:2017ynq}. Observationally, only Bo\"{o}tes~III and Tucana~III are currently identified to have pericentre distances within $R \leq 10$ kpc \citep{Battaglia2022AA657A}. This is consistent with the prediction that subhalo-induced heating is important only in the outer disc regions $R\gtrsim 10$~kpc (Figs.~\ref{fig:Heating_Decomp},~\ref{fig:Age_Weighted_Profiles}). 

On the scales of $\bigO(\rc)$, the tidal mass loss behaviour becomes less predictive. For initially unperturbed solitons on a circular orbit, \citet{Du:2018qor} observe that complete core disruption occurs within one orbital time if $\rt \lesssim \rc$. However, accreted subhalo orbits are typically elliptical with large apo-to-pericentre ratios \citep[e.g.][]{Ghigna1998MNRAS300,Jiang2015MNRAS448,Battaglia2022AA657A}. It is thus unclear  whether the same conclusion remains unchanged over a wider range of realistic host-subhalo setups, especially in the regime where impulsive tidal shocks dominate \citep{vandenBosch:2017ynq}, and convergent for higher simulation resolutions \citep{vandenBosch2018MNRAS475}. In light of these uncertainties, we have assumed in this work that instantaneous complete FDM subhalo disruption occurs if $\rt \leq 3\rc$, which likely overestimates the tidal effect on soliton cores and yields a more conservative FDM constraint. To fully assess the impact of uncertainties in the core-halo relation and FDM SHMF, numerical treatments are necessary to also satisfactorily address other factors like infall time and orbital eccentricity.

\subsubsection{Observational Inputs and Stellar Heating Estimate}\label{sssec:Obs_Input_Stellar_Heating}

The precise virial mass of the Galactic DM halo is still uncertain by a factor of 4 (see Sec.~\ref{ssec:MW_Background_Disc_sigma_z}), and $\sigmah$ cannot be directly constrained. The local DM density inferences show a similar degree of uncertainty $\rho_\text{DM}(R_\odot) \simeq0.006$ \textendash$0.018$ M$_\odot$pc$^{-3}$ (e.g. \citealp{Casagrande2020ApJ89626C, deSalas2021RPPh84j4901D} and references therein). The Galactic DM halo profile adopted in this work (yellow curves in Fig.~\ref{fig:MW-Total}) has $\rhoh(R_\odot) = 0.0080$~M$_\odot$pc$^{-3}$, indicating that the predicted disc heating rate could be an underestimation. Indeed, repeating the same analysis in Sec.~\ref{sssec:Solar_Neighbourhood} with now $\rhoh(R_\odot) = 0.018$ M$_\odot$pc$^{-3}$ pushes the exclusion bound derived from the observed $\sigma_z(R_\odot,\tage)$ to $m_a \gtrsim 0.7\times 10^{-22}$~eV. On the other hand, the cuspy nature of an NFW profile can overestimate the central FDM density and hence the FDM-driven stellar heating in the inner disc region $R \lesssim 3$~kpc.

Inference of the initial MW mass distribution also introduces an additional factor of uncertainty. The Galactic disc still lacks a tightly constrained mass accretion history (e.g. bottom panel of Fig. \ref{fig:MW-Disc_Mass_Infall}) and radial migration model. The reconstructed MW baryon structure evolution (Sec.~\ref{ssec:Disc_Mass_Accretion}) is expected to be less reliable at sufficiently high redshift $\redz \gtrsim 2$. In this work we consider a redshift-independent background density distribution $\rho_\text{bg}$. The DM halo has both $\rhoh$ and $\sigma_z$ decreasing with increasing redshift. However given the strong $\rhoh$- and $\sigma_z$-dependence of the heating rate $\bigH \propto T_\text{heat}^{-1} \propto \sigmah^{-6}\rhoh^2$, adopting an observationally ill-constrained $\redz$-dependent DM profile would introduce additional sizeable modelling uncertainty. 

The analytical estimate of stellar heating rate \eref{eqn:Transient_Perturbation_Rate} assumes a homogeneous background of identical perturbers, which roughly holds only as a local approximation. The typical uncertainty associated with precise definitions of the Coulomb factor $b_\text{max}/b_\text{min}$ is about an order of magnitude \citep[e.g.][]{Binney&Tremaine2008}. Furthermore, the FDM halo granule effective mass as well as the physical size and mass of orbiting subhaloes are strongly dependent on $R$ in the MW inner halo. Given the adiabatic cutoff length scale $b_\text{max} \simeq 5$~kpc at $R_\odot$, the relevant impact parameters sampled by disc stars encompass a largely inhomogeneous background of perturbers. For $m_a = 0.3$\textendash$5.0\times10^{-22}$~eV, the value of $\ln\Lambda\simeq 4$ for granule-induced (subhalo-induced for $R \gtrsim 14$ kpc) heating can be susceptible to the sizeable ambiguity in the Coulomb logarithm definition. 

The analytical framework presented in Sec. \ref{ssec:FDM_Halo_Granulation} assumes an FDM-only background with a Gaussian-distributed $\sigmah$. However the MW mass distribution becomes baryon-dominated for $R \lesssim 10$ kpc (see Fig. \ref{fig:MW-Total}). Since baryon corrections to the granulation size $b_\text{min}$ could to be important near the Galactic centre \citep{Chan2018MNRAS, Bar:2019bqz,Veltmaat:2019hou}, we expect the quoted halo granule attributes including $M_\text{gra}, \sigmah, b_\text{min}$ as well as the predicted heating behaviour in the inner disc region to somewhat deviate from the present calculations. The level of discrepancy will have be probed with self-consistent FDM+baryon simulations.

\subsubsection{Comparison with \citetalias{Church2019MNRAS}}\label{sssec:Comparison_with_C2019}

Our previous discussion, \citetalias{Church2019MNRAS}, arrives at the exclusion bound $m_a > 0.6\times 10^{-22}$~eV by requiring that the predicted stellar heating over $12$ Gyr cannot exceed $\sigma_z(R_\odot) = 32$ km s$^{-1}$. A similar consideration requiring the age-distribution-weighted $\sigma_z(R_\odot) \leq 22$ km s$^{-1}$ gives $m_a \gtrsim 0.4\times 10^{-22}$ eV in our analysis. We however caution that these constraints should not be taken at face value, in the presence of several sources of significant error and uncertainties detailed in Secs. \ref{sssec:Core-halo_Soliton_Size}, \ref{sssec:SHMF_Tidal}, and \ref{sssec:Obs_Input_Stellar_Heating}.

In addition to the aforesaid caveats, here we summarise the main differences and necessary changes made between the two works. The FDM-induced heating efficiency scales as $T_\text{heat}^{-1} \propto \Mgra^2\propto \sigmah^{-6}\rhoh^2$.  At leading order, \citetalias{Church2019MNRAS} overestimate the granule mass $\Mgra$ defined in Eqs. (18, 23) therein by a factor of $\simeq 19$; cf. \eref{eqn:M_gra} and footnote~\ref{fn:Halo_Velocity_Dispersion}. Relatedly their adopted $\sigmah = 200$~km~s$^{-1}$ (cf. $\simeq 86$~km~s$^{-1}$ extracted from FDM simulations) is noticeably higher than realistic values for an MW-sized halo inferred from Jeans modelling. \citetalias{Church2019MNRAS} adopt the CDM subhalo truncation prescription for FDM subhaloes, neglecting the more concentrated mass redistribution due to the presence of central solitons \eref{eqn:Tidal_chi}. Furthermore, the explicit redshift-dependences in both the FDM SHMF and disc surface density are not accounted for by \citetalias{Church2019MNRAS} in evaluating the adiabatic cutoffs $b_\text{max}(\redz)$. We also (1) consider other Galactic non-disc components and the BGD limit, (2) examine the effect of stellar age distribution on $\sigma_z$, (3) include the missing factor of $I_P(\zmax/h)$ in the Coulomb logarithm of subhalo-induced heating, cf. Eq. (17) of \citetalias{Church2019MNRAS}, and (4) provide an observationally constrained baryon mass infall model (Fig. \ref{fig:MW-Disc_Mass_Infall}).

Despite these significant changes in the detailed treatment, the overall conclusion remains the same. With a particle of $m_a \sim 0.5\times 10^{-22}$~eV, we can reproduce the scale height distribution and formation of the thick disc of the MW.

\subsection{Compatibility with Other FDM Constraints}\label{ssec:Literature_Comparison}

The light FDM mass scale $m_a = 0.5$\textendash$0.7\times 10^{-22}$ eV favoured by the observed Galactic disc thickening appears in tension with some recent exclusion bounds. For instance, the Lyman-$\alpha$ forest flux power spectrum has been exploited to place strong mass constraints $m_a \gtrsim0.7$\textendash$20 \times 10^{-21}$eV, with precise exclusion bounds depending strongly on the adopted modelling of the intergalactic medium and FDM suppression effect \citep{Kobayashi:2017jcf,Irsic:2017yje,Armengaud:2017nkf,Nori:2018pka,Rogers:2020ltq}. However the validity of these constraints is still greatly debated owing to a myriad of astrophysical modelling assumptions and data interpretation uncertainties \citep{Hui:2016ltb, Garzilli:2019qki,Bar-Or2019ApJ, Schutz:2020jox, Pozo:2020ukk}. In particular, the conventional interpretation of the Lyman-$\alpha$ forest neglects the structure induced by radiative transfer effects. 

Here we point out that a similar degree of theoretical and systematic uncertainties discussed in Sec.~\ref{ssec:Sources_of_Error} is also present in the FDM mass constraints derived from the empirical core-halo relation in Sec.~\ref{ssec:FDM_Core-Halo_Relation} and the FDM SHMF in Sec.~\ref{ssec:FDM_SHMF_Issues}.

\subsubsection{Core-halo Relation and Soliton Density Profiles}\label{ssec:FDM_Core-Halo_Relation}

Comparing the observed near-constant density cores in DM-dominated dwarf galaxies and FDM soliton profiles is found to favour $m_a\sim 10^{-22}$~eV \citep[e.g.][]{Schive:2014dra, Marsh:2015wka,Calabrese:2016hmp,Chen:2016unw,Wasserman2019ApJ885} or yield exclusion bounds $m_a \gtrsim 10^{-21}$\textendash$10^{-20}$~eV (\citealp[e.g.][]{Bernal2018MNRAS475, Hayashi:2021xxu, Bar:2021kti, Goldstein:2022pxu}; see however \citealp{Khelashvili:2022ffq}). The \textit{unperturbed} ground-state soliton density distribution with a core density $\rhoc$ has been independent verified by e.g. \citet{Schive:2014dra, Schwabe:2016rze,Mocz:2017wlg, May:2021wwp, Chan:2021bja} to follow
\begin{eqnarray}\label{eqn:Soliton_Den_Profile}
	\rho_\text{sol}(r) =\rhoc\bigg[1+9.1\times 10^{-2}\bigg(\frac{r}{\rc}\bigg)^2\bigg]^{-8},
\end{eqnarray}
Soliton cores formed in self-consistent merger and cosmological simulations additionally exhibit order-unity density oscillations on the timescale \citep{Veltmaat:2018dfz, Chiang:2021uvt}
\begin{eqnarray}\label{eqn:Soliton_Osc_Period}
	\tau_\text{sol}(\rhoc) = 92.1 \bigg(\frac{\rhoc}{\text{M}_\odot \text{pc}^{-3}}\bigg)^{-1/2} \text{ Myr},
\end{eqnarray}
that can generate adiabatic fluctuations in the background gravitational potential imprinted in the stellar kinematical data. Perturbed solitons have non-zero excited-state superposition coefficients and deviate from the analytical ground-state profile \eref{eqn:Soliton_Den_Profile}, increasing with density oscillation amplitude \citep{Li:2020ryg, Zagorac:2021qxq}. These perturbed configurations, albeit neglected in most analytical analyses, are ubiquitous in a cosmological snapshot by noting that a core DM density of $\bigO(1) \text{ M}_\odot \text{pc}^{-3}$ corresponds to $\tau_\text{sol} = \bigO(0.1)$~Gyr. 

More importantly, constraints of this type assume a one-to-one mapping between $m_a$ and soliton profiles via $M_\text{sol} \propto \Mh^{1/3}$ \eref{eqn:Core_Halo_Relation} with no statistical spread. However, a factor-two statistical uncertainty as well as other steeper power law indices $M_\text{sol} \propto \Mh^\alpha$ with $\alpha > 1/3$ have been reported (see Sec. \ref{sssec:Core-halo_Soliton_Size}). Larger $\alpha$ would give rise to smaller soliton core radius $\rc$ for fixed $m_a$ and $\Mh$. Furthermore existing numerical investigations into the $M_\text{sol}$\textendash$\Mh$ correlation have satisfactorily explored only $m_a \simeq 10^{-22}$~eV. It is unclear whether considerable extrapolation of the empirical core-halo relation to heavier $m_a\gg 10^{-22}$~eV still yields reliable inferences.

\subsubsection{FDM Subhalo Mass Function}\label{ssec:FDM_SHMF_Issues}

Several recent works \citep{Marsh:2018zyw, Banik:2019smi, Nadler:2019zrb, DES:2020fxi, Benito:2020avv, Schutz:2020jox} resort to the suppressed low-mass FDM subhalo abundance, or heuristically mapping between warm DM and FDM observables, and place exclusion bounds $m_a \geq 2$\textendash$5\times 10^{-21}$~eV. As highlighted in Sec.~\ref{sssec:SHMF_Tidal}, various independent approaches all yield conflicting FDM SHMFs, except for the simulations evolved from FDM initial conditions by \citet{Schive:2015kza} and \citet{May2022arXiv220914886M} compared at $\redz = 3$. The discrepancy is especially pronounced in the low-mass tail. 

These constraints derived from the low-mass tail of subhalo distribution can also be sensitive to the adopted tidal stripping prescription and largely unknown mass-concentration relation of FDM subhaloes. Statistical spread and uncertainties in the core-halo relation (Sec.~\ref{sssec:Core-halo_Soliton_Size}) would further enhance the halo-to-halo variance in the post-accretion density profiles of low-mass FDM subhalo populations. For instance, varying the adopted warm DM mass-concentration relations, albeit found in \citet{Du:2016zcv} to be inconsequential to the resulting FDM SHMF, requires further numerical corroboration, as the leading-order discrepancy amongst analytical and empirical FDM subhalo distributions is already sizeable (Fig. \ref{fig:FDM_SHMF}).

We caution that at least some preliminary convergence should be reached before meaningful particle bounds could be inferred from FDM SHMF. Cosmological simulations of \citet{Schive:2015kza} and \citet{May2022arXiv220914886M} have respectively probed up to only $m_a \lesssim 3.2(0.7)\times 10^{-22}$~eV and $\redz \gtrsim 3$. Given the lack of fully self-consistent FDM cosmological simulations reaching heavier particle masses $m_a \gg 10^{-22}$~eV and lower redshifts, the validity of extrapolating SHMF to $m_a \gtrsim 10^{-21}$~eV and $\redz \simeq 0$ remains to be verified and could undermine the robustness of constraints derived from a given fiducial SHMF profile. Here we include some brief remarks:
\begin{itemize}
	\item \citet{Marsh:2018zyw} integrate the FDM SHMF derived in \citet{Du:2016zcv} over the measured $\pm2\sigma$ mass range $0.4$\textendash$2\times 10^7$~M$_\odot$ of Eridanus II  to exclude $m_a \gtrsim 8\times 10^{-22}$ eV. The constraint relies on the largely uncertain low-mass tail; see also \citet{Chiang:2021uvt}.
	\item \citet{Nadler:2019zrb} adopt a direct one-to-one mapping of the half-mode scales between FDM and thermal relic warm DM transfer functions, and derive $m_a \geq 2.9\times 10^{-21}$~eV from the observed MW satellite population. Disregarding detailed features of FDM SHMF, this simplified translation in the quasilinear regime has been argued to be inaccurate \citep{Irsic:2017yje,Schutz:2020jox,DES:2020fxi}.
	\item \citet{Schutz:2020jox} finds $m_a \gtrsim 2\times 10^{-21}$ eV via bounding the FDM SHMF of \citet{Du:2016zcv} by the already excluded warm DM SHMFs from quasar lensing and stellar streams. The analysis is sensitive to the ill-constrained subhalo $\leq \bigO(10^8)$~M$_\odot$ distribution.
	\item \citet{Benito:2020avv} adopt the FDM SHMF of \citet{Schive:2015kza} and an analytical subhalo mass loss model. The constraint $m_a \geq 5.2\times 10^{-21}$ eV is inferred from the possible presence of low-mass subhaloes $10^{7\text{\textendash}9}$~M$_\odot$ given the perturbed stellar streams.
	\item \citet{Banik:2019smi} rely on the semi-analytical FDM SHMF of \citet{Du:2016zcv}; the dark substructure mass inferred from perturbed stellar stream is used to set $m_a \gtrsim 2\times 10^{-21}$ eV.
	\item \citet{DES:2020fxi} adopt the semi-analytical prescription of \citet{Du:2016zcv} and \eref{eqn:M0_FDM_SHMF_Schive} to place the constraint $m_a \geq 2.9\times 10^{-21}$~eV from two simulated MW-like satellite populations.
\end{itemize}

\subsection{Formation of Ultra-thin Disc Galaxies via Anisotropic-heating-induced Radial Migration}\label{ssec:Anisotropic_Heating}

The existence of ultra-thin (or super-thin) disc galaxies has long been established observationally \citep[e.g.][]{Goad1981ApJ250,Matthews1999AJ118}, yet the physical origin remains puzzling \citep[e.g.][]{Kautsch2009PASP121,Banerjee2013MNRAS431}. The observed small vertical-to-planar scale length ratios $h/R_0\leq 0.1$ translate to a scale separation in the vertical and azimuthal adiabatic cutoffs. This non-trivial implication prompts the intriguing exploration of ultra-thin morphology as an attractor solution of the FDM-substructure-induced heating.

For an FDM halo with approximately isotropic velocity dispersion, the level of disc heating anisotropy along the $\hat{z}$- and $\hat{\phi}$-direction correlates with the respective Coulomb logarithms (e.g. see footnote \ref{fn:Coulomb_Logarithm}) that in turn depend on the vertical and azimuthal oscillation timescales $t_z$ and $t_\phi$. Roughly speaking when the ratio
\begin{eqnarray}\label{eqn:Adiabatic_Time_Scales}
	\frac{t_z}{t_\phi} \eee \frac{h/\sigma_z}{R_0/\vc}
\end{eqnarray}
is sub-unity, vertical heating becomes less effective due to a smaller adiabatic cutoff  compared to the $\hat\phi$-direction counterpart. Azimuthal heating propels disc stars to undergo outward radial migration. In the limit $t_z/t_\phi \ll 1$, the initial disc scale height $h$ is preserved by adiabatic invariance, whilst the strong radial migration naturally elongates the scale radius $R_0$ over time, resulting in $h/R_0 \rightarrow 0$. 

\begin{figure} 
	\centering	
	\includegraphics[width=0.98\linewidth]{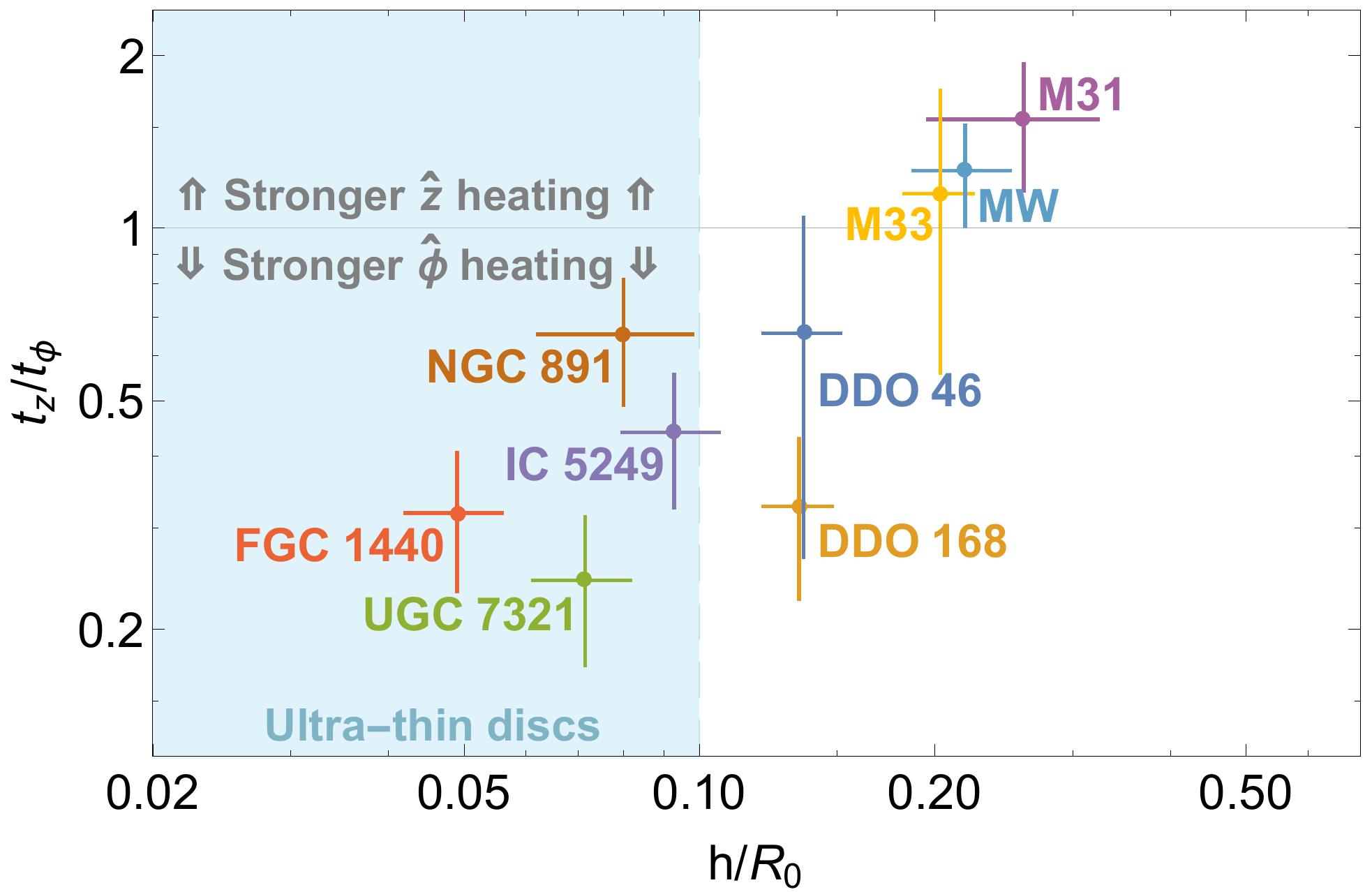}
	\caption{Inferred ratio of vertical to azimuthal oscillation timescales $t_z/t_\phi$ as a function of the observed ratio between disc scale height and length $h/R_0$; see text for error bar estimates. The correlation appears consistent with the interpretation that ultra-thin disc galaxies could serve as an attractor solution to the strongly anisotropic heating induced by FDM substructures.}
	\label{Anisotropic_Heating}
\end{figure}

To investigate whether such a correlation indeed exists between the ultra-thin morphology $h/R_0$ and scale-separated dynamical imprint $t_z/t_\phi$, we compile in Fig. \ref{Anisotropic_Heating} a list of disc galaxies with varying $h/R_0$. The vertical velocity dispersion $\sigma_z$ and circular rotation velocity $\vc$ are taken at the radius $R_0$. For the MW, a fiducial disc height $h_z = 0.5$~kpc (average of thick and thin discs) is adopted. M31 thin and thick disc components are averaged to give $h = 2.0\pm 0.4$~kpc, $R = 7.7\pm1.1$ kpc, $\sigma_z = 43\pm1.5$ km s$^{-1}$, and $\vc = 256 \pm 4$ km~s$^{-1}$ \citep{Collins2011MNRAS413, Sofue1999ApJ}. M33 has $h \simeq 0.37$ kpc, $R = 1.82\pm0.02$~kpc, $\vc = 79\pm4$  km s$^{-1}$, and $\sigma_z = 14\pm7$ km s$^{-1}$ \citep{Kam2015MNRAS449,Ciardullo2004ApJ614}. A relative standard deviation of $0.1$ is adopted when the error bar associated with any quoted measurement is unavailable, which in most cases is likely an underestimate. The overall trend is consistent with the explanation that ultra-thin discs \mbox{arise from strongly anisotropic FDM-induced heating.} 

As many thin disc dwarf galaxies of interest still lack spectral measurements of $\sigma_z$, here we caution the robustness of indirect inferences and list some pertinent caveats: \textbf{DDO 46} with $h =155$~pc, $R_0 = 1.14\pm 0.06$ kpc has only one direct measurement of $\sigma_z = 13.5\pm8$ km s$^{-1}$ \citep{Hunter2012AJ144,Johnson2015AJ149}. \textbf{DDO~168} with $h = 110$~pc, $R_0 = 0.82 \pm 0.01$~kpc, and $\rhoc \simeq 10^{-2}$~M$_\odot$pc$^{-3}$ is baryon-dominated within $R_0$ where $\sigma_z = 10.7\pm2.9$ km s$^{-1}$ is observationally constrained \citep{Hunter2012AJ144,Johnson2015AJ149,Oh2015AJ149}; the effect of FDM-induced heating in the inner disc is thus expected sub-dominant. Otherwise the thin vertical height is preserved by adiabatic invariance $t_z \simeq 10$~Myr $\ll\tau_\text{sol} \simeq 920$ Myr. \textbf{UGC 7321} with $h = 150$~pc, $R_0 = 2.1$~kpc has $\sigma_z = 20.4$~km~s$^{-1}$ indirectly inferred from the disc luminosity profile \citep{Matthews1999AJ118,Matthews2000AJ120}. \textbf{FGC 1440} with $R_0/h = 20.4$ has the ratio $v_c/\sigma_z = 6.5\pm 1.5$ inferred indirectly from Jeans modelling \citep{Karachentsev1999BSAO47,Aditya2022MNRAS509}. \textbf{IC 5249} with $h = 0.65$~kpc, $R = 7$~kpc, $v_c = 90\pm5$~km~s$^{-1}$ has $\sigma_z = 19\pm 4$~km~s$^{-1}$ derived from the disc surface density \citep{vanderKruit2001AA379}. \textbf{NGC 891} with $h = 0.4$~kpc, $R = 5\pm1$~kpc, $v_c = 220$~km~s$^{-1}$ has $\sigma_z = 27$~km~s$^{-1}$ derived from the \mbox{luminosity profile \citep{Oosterloo2007AJ134,Yim2014AJ148}.}

\section{Summary and Conclusions}\label{sec:Conclusions}

The present-day Galactic disc morphology provides a unique observational handle on the nature of early-time  dynamical heating mechanisms. We have provided a revised, more detailed treatment of disc stellar heating caused by FDM substructures, building upon the analytical framework developed in  \citetalias{Church2019MNRAS}. The main improvements include: (1) accounting for the full MW baryon and DM distributions, (2) modelling the disc profile and dynamics in the BGD limit, (3) adopting $\sigma_z$ from recent large-scale kinematic measurements and $\sigmah$ from self-consistent FDM simulations, (4) sharpening the analytical estimate of stellar heating induced by FDM halo granulation and subhalo passages, (6) calibrating the baryon infall model with observational inferences, (7) tracking the distribution of stellar ages, and (8) computing the age-distribution-weighted $\sigma_z$ profiles.

The heating mechanism predicts a flexible age\textendash velocity dispersion relation in the solar vicinity $\sigma_z\propto t^\beta$ with $\beta = 0.40$\textendash$0.20$ for $m_a = 0.3$\textendash$1.0\times 10^{-22}$ eV. A lower bound $m_a \gtrsim 0.4\times 10^{-22}$~eV can be derived from the observed stellar kinematics measurements of low-[$\alpha$/Fe] disc star populations. Substantial gravitational heating caused by halo granulation and orbiting subhaloes offers a compelling explanation for the observed inner disc thickening and outer flaring of young stellar populations, favouring $m_a \simeq  0.5$\textendash$0.7\times 10^{-22}$ eV. Even for heavier $m_a$, this non-uniform heating signature remains consistent with the observed U-shaped $\sigma_z(R)$ profile that has the global minimum of disc heating rate located near the solar vicinity.

The FDM substructure-driven disc heating asymptotes the CDM prediction for heavier $m_a$, and becomes largely indistinguishable beyond $m_a \gtrsim 5\times10^{-22}$~eV. Hence if sufficient disc thickening mechanisms already exist in the canonical CDM paradigm, FDM will still not be entirely ruled out by the observed age\textendash velocity dispersion relation; instead this particular astrophysical observable is sensitive only up to the mass scale $m_a\sim5\times10^{-22}$~eV. We further point out that the dynamical impact of Gaia\textendash Enceladus and Sgr dwarf mergers alone, albeit important in shaping the Galactic stellar populations, is likely insufficient to account for the observed thick disc heating in the solar neighbourhood.
	
We also examine non-trivial modelling uncertainties present in the empirical core-halo relation, FDM SHFM, FDM-driven stellar heating calculations, and observational inputs. In particular, the ill-constrained (inner) DM density profile, uncertain subhalo mass loss due to Galactic tides, and efficient radial migration prevent us from placing robust $m_a$ constraints from the observed $\sigma_z$ profiles.

Other existing FDM constraints $m_a \gtrsim 10^{-20}$\textendash$10^{-21}$~eV derived from Lyman-$\alpha$ forest, cored density profile fitting, stellar streams, and the MW satellite populations are in tension with the particle mass $m_a \simeq  0.5$\textendash$0.7\times 10^{-22}$ eV favoured by the local age\textendash velocity dispersion relation. As shown in Secs.~\ref{ssec:Sources_of_Error} and \ref{ssec:Literature_Comparison}, the FDM observables quoted in many of these analyses are over simplified (e.g. neglecting intrinsic statistical scatters) or fairly uncertain (e.g. lacking numerical corroborations in the $m_a$ range of interest). The robustness of extrapolating the empirical core-halo relation and FDM SHMF well beyond the numerically verified regime $\bigO(10^{-22})$~eV (and down to low redshifts for the latter) is currently unknown and awaits verification from future simulations. We point out that some existing exclusion bounds on $m_a$ could be significantly relaxed as a result.

Given the heating efficiency scaling as $T_\text{heat}^{-1} \propto \sigmah^{-6}\rhoh^2$, the FDM-induced heating becomes increasingly pronounced near the galactic centre (assuming a standard NFW profile). For the MW, the baryon-dominated innermost region $R\lesssim 2$ kpc makes the robust inference of disc heating and observational constraints on $\sigma_z(R)$ challenging. In contrast, bulgeless disc dwarf galaxies that are DM-dominated down to small radii are expected to yield competitive $m_a$ constraints derived from stellar heating. The robustness again relies on the direct measurements of $\sigma_z(R)$ as well as convergent inferences about the disc mass profile, robust stellar age information, and DM density and velocity dispersion profiles. 

For sufficiently small disc scale heights, background potential perturbations become adiabatic with respective to the rapid vertical oscillations of disc stars, suppressing the $z$-direction stellar heating. As a consequence, the formation of ultra-thin disc galaxies naturally arises as an attractor solution to the strongly anisotropic stellar heating caused by FDM density perturbations.

Similar to the spin-0 (scalar) ultralight DM considered in this work, theoretically well-motivated spin-1 (vector) ultralight DM \citep[e.g.][]{Agrawal:2018vin} also gives rise to haloes exhibiting wave interference patterns \citep{Amin:2022pzv}. These granular features of vector DM can drive stellar disc thickening in a qualitatively similar fashion.

\section*{Acknowledgements}
We are happy to acknowledge useful conversations with Simon May and constructive comments from the anonymous referee. We thank Melissa Ness for the helpful discussions on the current observational status of relevant large spectroscopic surveys and recent literature on stellar kinematics of the Galactic disc. We also thank Guan-Ming Su for providing the Milky-Way-like dark matter density and velocity dispersion profiles extracted from self-consistent fuzzy dark matter haloes. We acknowledge Lam Hui for the illuminating insights on the interplay between FDM-induced heating and ultra-thin/bulgeless disc galaxies. HS acknowledges funding support from the Jade Mountain Young Scholar Award No. NTU-111V1201-5, sponsored by the Ministry of Education, Taiwan. This research is partially supported by the National Science and Technology Council (NSTC) of Taiwan under Grants No. NSTC 111-2628-M-002-005-MY4 and No. NSTC 108-2112-M-002-023-MY3, and the NTU Academic Research-Career Development Project under Grant No. NTU-CDP-111L7779.


\section*{Data Availability}

The data underlying this article will be shared on reasonable request to the corresponding author.


\bibliographystyle{mnras}
\bibliography{MyBibTeX1}


\appendix

\section{Impulsive Heating to Adiabatic Invariants}
\label{app:disc_heating_derivation}

Appendices \ref{app:General_Heat_Eq}, \ref{app:SGD_Limit} review the derivation of the SGD-limit heating master equation first presented in \citetalias{Church2019MNRAS}. In Appendix \ref{app:BGD_Limit}, we compute the suitable expression in the BGD limit. 

\subsection{Generic Heating Equation}\label{app:General_Heat_Eq}

The vertical action $J_z(R,\zmax) \eee 2m\oint p dq$ \footnote{The convention adopted here is consistent with \citetalias{Church2019MNRAS} (see Appendix A therein) and differs from the usual definition $J\eee \frac{1}{2\pi}\oint \mathbf{p}\cdot d\mathbf{q}$ \citep{Binney&Tremaine2008}, where $\mathbf{p}$ and $\mathbf{q}$ are the canonical coordinates of interest.} of a locally isothermal disc with negligible radial migration can be expressed as
\begin{eqnarray}\label{app:Jz_P}
	\begin{cases}
		J_z(R,\zmax) =  2m\int_{-\zmax}^{\zmax} dz' \sqrt{2\big[\phi_\text{tot}(R,\zmax)-\phi_\text{tot}(R,z')\big]},\\
		E(R,\zmax) = \frac{p_z^2}{2m} + m\phi_\text{tot}(R,z) = m\phi_\text{tot}(R,\zmax),\\
		P(R,\zmax) \eee \frac{\pd J_z}{\pd E} = 2\int_{-\zmax}^{\zmax} dz' \frac{1}{\sqrt{2[\phi_\text{tot}(R,\zmax)-\phi_\text{tot}(R,z')]}},
	\end{cases}
\end{eqnarray}
where $E(R,\zmax)$ denotes the Hamiltonian for a disc star of mass $m$ undergoing vertical oscillations reaching $\zmax$ at a radius $R$. Under a time-varying potential parameterised by $\lambda_i$, transient perturbations induce irreversible energy transfer at a rate given by \eref{eqn:Transient_Perturbation_Rate}
\begin{eqnarray}\label{app:Impulsive_Energy_Injection}
	\frac{\pd E}{\pd t} = \frac{m}{2}\frac{d \sigma_z^2}{dt}\bigg|_\text{impulsive}  \eee  \frac{m}{2}\bigM\log\bigg(\frac{P}{\tau}\bigg),
\end{eqnarray}
that drives the time evolution of vertical action
\begin{eqnarray}\label{app:Jz_Time_Evolution}
	\frac{d J_z}{dt} = \frac{\pd J_z}{\pd E}\bigg( \frac{d E}{d t}\bigg) + \frac{\pd J_z}{\pd \lambda_i} \frac{d \lambda_i}{ dt}= \bigg( P\frac{\pd E}{\pd \lambda_i}+ \frac{\pd J_z}{\pd \lambda_i}\bigg)  \frac{d \lambda_i}{ dt} + P\frac{\pd E}{\pd t},
\end{eqnarray}
 where $\bigM$ and $b_\text{min}$ depend only on the perturber properties. If $\lambda_i$ vary slowly compared to the orbital timescale, then the adiabatic invariance of $J_z$ implies
\begin{eqnarray}
	P \frac{\pd E}{\pd \lambda_i} + \frac{\pd J_z}{\pd \lambda_i} = P \frac{\pd E}{\pd \lambda_i} - P\bigg\<\frac{\pd E}{\pd \lambda_i}\bigg\>_\text{orbit} \simeq 0,
\end{eqnarray}
and \eref{app:Jz_Time_Evolution} reduces to $d J_z/dt  = P(\pd E/\pd t)$, or equivalently
\begin{eqnarray}\label{app:Heat_Eq}
	\frac{d (J_z/2m)}{ dt } = \frac{P\bigM}{4}\log\bigg(\frac{P}{\tau}\bigg).
\end{eqnarray}

The present derivation assumes only that disc heating is sourced by impulsive, weak gravitational encounters. The master equation \eref{app:Heat_Eq} can hence model the heating rate of any analytical or numerical disc profile with adiabatic invariant $J_z$ induced by (e.g.) DM subhalo passages and FDM halo granulation. Note that the master equation carries explicit $\zmax$-dependence via $J_z(R,\zmax)$ and $P(R,\zmax)$, which can be integrated out if the vertical number density distribution of disc stars is known a priori. We focus on the SGD (Appendix \ref{app:SGD_Limit}) and BGD limits (Appendix \ref{app:BGD_Limit}) where the disc profiles are analytical \eref{eqn:Disc_Den_Profiles}, and derive the corresponding $\zmax$-independent, ensemble-averaged heat equations.

\subsection{Self-gravity-dominated (SGD) Limit}\label{app:SGD_Limit}

Given the analytical disc potential and vertical oscillation period \pCAeref{eqn:Disc_Gra_Potential}\pCBeref{eqn:Vertical_Potentials}\pCCeref{eqn:Disc_Osc_Period} in the SGD limit, the corresponding vertical action as defined in \eref{app:Jz_P} can be expressed as
\begin{eqnarray}\label{app:Jz_P_Self_Gravity_Dom}
	\begin{cases}
		\frac{J_z}{2m} = \frac{2\sigma_z^3}{\pi G \Sigma} I_{J1}(\zmax/h),\\
		I_{J1}(x) \eee \int_{-x}^x dx'\sqrt{\log(\cosh x/\cosh x')}.
	\end{cases}
\end{eqnarray}
The master equation \eref{app:Heat_Eq} describing the time evolution of $J_z$ of a single disc star indexed by $(R,\zmax)$ reduces to
\begin{eqnarray}\label{app:Single_Star_Action}
	\scalebox{0.95}{\text{$\frac{3}{2}\Big(\frac{d\sigma_z^2}{dt} \Big)I_{J1} -\frac{\sigma_z^2}{\Sigma}\Big(\frac{d\Sigma}{dt} \Big) I_{J1} + \sigma_z^2\Big(\frac{dI_{J1}}{dt}\Big)= \frac{\bigM}{8}\Big[I _{P1}\log\Big(\frac{\sigma_zI_{P1} }{\pi G \Sigma \tau}\Big)\Big].$}}
\end{eqnarray}
We are interested in the \textit{ensemble-averaged}, $\zmax$-independent disc heating rate at each radius $R$. The local isothermality underpinning \pCAeref{eqn:Hydrostatic_Poisson}\pCCeref{eqn:Disc_Den_Profiles} implies that disc stars labelled by $x\eee z_\text{max}/h$ are distributed in accord with $\text{sech}^2(x)$. With
\begin{eqnarray}
	\begin{cases}
		\<I_{J1}\> = \int_0^\infty dxI_{J1}(x)\text{sech}^2(x) \simeq 0.759,\\
		\big\<\frac{dI_{J1}}{dt}\big\> = \frac{1}{N}\sum\limits_{\scriptscriptstyle{i=1}}^{\scriptscriptstyle{N}} \frac{dI_{J1}(x_i)}{dt}= \frac{d}{dt}\Big(\frac{1}{N}\sum\limits_{\scriptscriptstyle{i=1}}^{\scriptscriptstyle{N}} I_{J1}\Big) = \frac{d\<I_{J1}(x)\>}{dt} = 0,\\
		\<I_{P1}\> = \int_0^\infty dxI_{P1}(x)\text{sech}^2(x) \simeq 4.79,\\
		\<I_{P1}\log(I_{P1})\> = \int_0^\infty dxI_{P1}(x)\log[I_{P1}(x)]\text{sech}^2(x) \simeq 7.53,
	\end{cases}
\end{eqnarray}
the distribution-averaged \eref{app:Single_Star_Action} in the SGD limit simplifies to
\begin{eqnarray}\label{app:Heating_Equation_Self_Gravity}
	\frac{d\sigma_z^2}{dt} =  \frac{2\sigma_z^2}{3}\frac{d\log\Sigma}{dt}+\kappa_1 \<\bigH\>_1,
\end{eqnarray}
where we have recovered Eq. (71) of \citetalias{Church2019MNRAS} and (cf. \pCAeref{eqn:Transient_Perturbation_Rate}\pCCeref{app:Impulsive_Energy_Injection})
\begin{eqnarray}\label{app:Ensem_Ave_Coefficients_Self_Gravity_Dom}
	\begin{cases}
		\kappa_1 \eee \frac{\<I _{P1}\>}{12\<I_{J1}\>}\simeq 0.526,\\
		\<\bigH\>_1  \eee \bigM\log(\<P\>_1/\tau),\\
		\<P\>_1 = \frac{\sigma_z I_{\Lambda1}}{\pi G \Sigma},\\
		I_{\Lambda1} \eee e^{\<I _{P1}\log(I_{P1})\>/\<I _{P1}\>} \simeq 4.82.
	\end{cases}
\end{eqnarray}
The factor $\kappa_1\simeq 0.5$ can be interpreted as a consequence of virial theorem, where the energy inflow is about equally distributed between the disc self-gravitation and kinetic energy.

\subsection{Background-dominated (BGD) Limit}\label{app:BGD_Limit}
To derive the master equation in the BGD limit, we first note that $\phi_\text{tot}(R,\Delta z)\simeq \phi_\text{bg}(R,\Delta z)$ has no closed-form expression. Given $\phi_\text{bg}(R,z)$ is approximately harmonic in $z$, to proceed analytically we assume $\phi_\text{bg}(R,z) \propto z^2$ exactly in this subsection. The ensemble-averaged vertical action \eref{app:Jz_P} $J_z \propto h^2$ is now independent of the disc surface density $\Sigma(R)$; see \eref{eqn:Scale_Height_Limits}. A similar derivation as that presented in Appendix \ref{app:SGD_Limit} gives the ensemble-averaged heating rate
\begin{eqnarray}\label{app:Heating_Eqaution_Backgroud_Dom}
	\frac{d \sigma_z^2}{dt} = \kappa_2\<\bigH\>_2,
\end{eqnarray}
where oscillations inside a harmonic potential $\<E\> = \sigma_z^2$ correspond to $\kappa_2 = 1$ \citep{Lacey1985ApJ, Binney&Tremaine2008}. Furthermore in a potential harmonic along the $z$ direction, the vertical oscillation timescale $P(R,\zmax) = P(R)$, or equivalently $2I_{P2}(R,\zmax) = 2I_{P2}(R)$ defined in \eref{eqn:Disc_Osc_Period}, is $\zmax$-independent. In the BGD limit, the vertical stellar distribution \eref{eqn:Disc_Den_Profiles} now follows $e^{-x^2}/(\sqrt{\pi}/2)$, where $x \eee \sqrt{\pi}\zmax/2h$. The ensemble-averaged heating rate is hence
\begin{eqnarray}\label{app:IP_2}
	\begin{cases}
		\<\bigH\>_2 \eee \bigM\log(\<P\>_2/\tau),\\
		\<I_{P2}\>(R) = \int_0^\infty dxI_{P2}\scalebox{0.85}{\Big(}R, \frac{2h}{\sqrt\pi}x\scalebox{0.85}{\Big)}\frac{e^{-x^2}}{\sqrt{\pi}/2} = I_{P2}(R,h(R)),\\
		\<P\>_2(R) = 2e^{\<I _{P2}\log(I_{P2})\>/\<I _{P2}\>} = 2I_{P2}(R,h(R)),
	\end{cases}
\end{eqnarray}
where the last equalities in the second and bottom lines follow directly from that $I_{P2}(R,z_1) = I_{P2}(R,z_2)$ for arbitrary $z_1, z_2 \in \mathbb{R}$ under a background potential harmonic in $z$.

\label{lastpage}
\end{document}